\documentclass[twocolumn,nofootinbib,notitlepage,amssymb]{revtex4-1}

\usepackage{amsmath}	
\usepackage{microtype}
\usepackage{graphicx}
\usepackage{color}
\usepackage{hyperref}
\usepackage{cleveref}

\makeatletter
\DeclareRobustCommand{\rcite}[1]{%
  \rcite@aux#1,\@nil{#1}%
}
\def\rcite@aux#1,#2\@nil#3{%
  \if\relax#2\relax
    Ref.~\cite{#3}%
  \else
    Refs.~\cite{#3}%
  \fi
}
\makeatother

\newcommand{\Mp}{M_\mathrm{Pl}}
\newcommand{\eff}{\mathrm{eff}}
\newcommand{\dd}{\mathrm{d}}
\newcommand{\sdg}{\sqrt{-g}}

\newcommand{\mn}{{\mu\nu}}

\newcommand{\AB}{{AB}}
\newcommand{\Ij}{{ij}}
\newcommand{\diag}{\operatorname{diag}}
\begin{document}

\title{Massive mimetic cosmology}

\author{Adam R. Solomon}
\email{adamsolo@andrew.cmu.edu}
\affiliation{Department of Physics \& McWilliams Center for Cosmology,\\Carnegie Mellon University, Pittsburgh, Pennsylvania 15213, USA}

\author{Valeri Vardanyan}
\email{vardanyan@lorentz.leidenuniv.nl}
\affiliation{Lorentz Institute for Theoretical Physics, Leiden University, P.O. Box 9506, 2300 RA Leiden, The Netherlands}
\affiliation{Leiden Observatory, Leiden University, P.O. Box 9513, 2300 RA Leiden, The Netherlands}

\author{Yashar Akrami}
\email{akrami@ens.fr}
\affiliation{D\'epartement de Physique, \'Ecole Normale Sup\'erieure, PSL Research University, CNRS, 24 rue Lhomond, 75005 Paris, France}
\affiliation{Lorentz Institute for Theoretical Physics, Leiden University, P.O. Box 9506, 2300 RA Leiden, The Netherlands}

%%%%%%%%%%%%%%%%%%%%%%%%%%%%%%%%%%%%%%%%%%%%%%%%%%%%%%%%%%%%
%TC:ignore
\begin{abstract} 
We study the first cosmological implications of the mimetic theory of massive gravity recently proposed by Chamseddine and Mukhanov. This is a novel theory of ghost-free massive gravity which additionally contains a mimetic dark matter component. In an echo of other modified gravity theories, there are self-accelerating solutions which contain a ghost instability. In the ghost-free region of parameter space, the effect of the graviton mass on the cosmic expansion history amounts to an effective negative cosmological constant, a radiation component, and a negative curvature term. This allows us to place constraints on the model parameters---the graviton mass and the St\"uckelberg vacuum expectation value---by insisting that the effective radiation and curvature terms be within observational bounds. The late-time acceleration must be accounted for by a separate positive cosmological constant or other dark energy sector. We impose further constraints at the level of perturbations by demanding linear stability. We comment on the possibility of distinguishing this theory from $\Lambda$CDM with current and future large-scale structure surveys.
\end{abstract}
%TC:endignore

\maketitle

%%%%%%%%%%%%%%%%%%%%%%%%%%%%%%%%%%%%%%%%%%%%%%%%%%%%%%%%%%%%
\section{Introduction}
\label{sec:intro}

Chamseddine and Mukhanov have recently proposed~\cite{Chamseddine:2018qym,Chamseddine:2018gqh} a novel ghost-free theory of massive gravity in which one of the four St\"uckelberg scalars is constrained in the same way as in the mimetic theory of dark matter~\cite{Chamseddine:2013kea}, spontaneously breaking Lorentz invariance. In this Letter, we study the immediate implications of this mimetic massive gravity for cosmological theory and observation.

From a field-theoretic perspective, general relativity is the unique theory (in four spacetime dimensions) of a massless spin-2 particle, or graviton. It is therefore natural to ask whether it is possible to endow the graviton with a non-zero mass, and what sort of theoretical structures would result \cite{Fierz:1939ix}. A closely related line of inquiry asks whether it is possible for two or more gravitons to interact \cite{Isham:1971gm}. Most nonlinear realizations of such theories suffer from the so-called Boulware-Deser ghost instability \cite{Boulware:1973my}. The past decade has seen the construction of models which avoid this instability, allowing for the construction of ghost-free theories of massive gravity \cite{Creminelli:2005qk,deRham:2010ik,deRham:2010kj,Hassan:2011hr,Hassan:2011tf,Hassan:2011ea,Hassan:2012qv} and bimetric and multimetric gravity \cite{Hassan:2011zd,Hassan:2011ea,Hinterbichler:2012cn}. We refer the reader to the reviews \cite{Hinterbichler:2011tt,deRham:2014zqa} on massive gravity and \cite{Solomon:2015hja,Schmidt-May:2015vnx} on bimetric gravity. The theory of mimetic massive gravity proposed in \rcite{Chamseddine:2018qym,Chamseddine:2018gqh} takes a new and alternative path to a ghost-free nonlinear theory of massive gravity.

A generic theory of massive gravity propagates six degrees of freedom, which should be thought of as the five helicity states of a massive graviton plus an additional, ghostly scalar. The easiest way to understand the degrees-of-freedom counting is to observe that a graviton mass breaks diffeomorphism invariance. This is a gauge symmetry and so can be restored by the addition of four St\"uckelberg scalars $\Phi^A$, which propagate in addition to the two (now potentially massive) tensor modes of general relativity.

As an illustration, consider a Lorentz-invariant theory of massive gravity. In order to construct non-trivial, non-derivative interactions for the metric, one requires a second ``reference" metric. The simplest choice for this metric is that of flat space, $\eta_\mn$, but the addition of this prior geometry breaks diffeomorphism invariance; for instance, there are preferred coordinate systems in which $\eta_\mn=\operatorname{diag}(-1,1,1,1)$. But diffeomorphism invariance is simply a redundancy in description, and can be restored by the addition of redundant variables, i.e., replacing $\eta_\mn \to \eta_{AB}\partial_\mu\Phi^A\partial_\nu\Phi^B$, where $\eta_{AB}=\operatorname{diag}(-1,1,1,1)$ and the four fields $\Phi^A$ transform as spacetime scalars. One can always, by means of a diffeomorphism, choose the unitary gauge in which $\Phi^A=x^A$, and we recover the original description of the theory in terms of a symmetry-breaking reference metric. Generic interaction terms for the graviton, e.g., generic functions of $g^{\mu\alpha}\eta_{AB}\partial_\alpha\Phi^A\partial_\nu\Phi^B$, will lead to dynamics for each of these four scalars, in addition to the two modes of general relativity, for a total of six degrees of freedom.

At the linear level, i.e., linearizing the metric about flat space in unitary gauge, $g_\mn=\eta_\mn+h_\mn$ and $\Phi^A=x^A$, we find that one of the six degrees of freedom leads to a ghost instability unless we specifically arrange the mass term into the Fierz-Pauli form, $\mathcal{L}_\mathrm{mass}\sim h_\mn^2-h^2$, in which case the dynamics of the ghostly mode take the form of a total derivative. Continuing this procedure at higher orders in perturbation theory---i.e., continually packaging ghostly operators into total derivative structures---leads uniquely to the non-linear massive gravity theory of de Rham, Gabadadze, and Tolley (dRGT) \cite{deRham:2010ik,deRham:2010kj}.

The recent proposal of Chamseddine and Mukhanov takes a novel alternative approach to eliminating the dangerous ghostly mode \cite{Chamseddine:2018qym,Chamseddine:2018gqh}. Noticing that the ghost can be associated to the $\Phi^0$ St\"uckelberg mode, they propose imposing the constraint $g^\mn \partial_\mu\Phi^0\partial_\nu\Phi^0=-1$. This is motivated by a similar construction known as mimetic gravity \cite{Chamseddine:2013kea}, in which the constrained scalar winds up behaving like dark matter.\footnote{For an earlier construction in which a constrained scalar mimics dark matter and dark energy, and which contains mimetic dark matter as a subset, see \rcite{Lim:2010yk}.} Mimetic massive gravity takes this constrained scalar to be one of the St\"uckelberg modes of a massive graviton, eliminating the ghost. They propose the following action, designed to ensure stability at the linear level (notice that the mass term is not of the Fierz-Pauli form),
\begin{align}
S &= \int\dd^4x\sdg\bigg[\frac{\Mp^2}{2}R+\frac{m^2\Mp^2}{8}\left(\frac12\bar h^2-\bar h_\AB^2\right) \nonumber \\
&\hphantom{{}=\int\dd^4x\sdg\bigg[} + \lambda(X+1)\bigg] + S_\mathrm{matter}, \label{eq:action}
\end{align}
with $X\equiv g^\mn\partial_\mu\Phi^0\partial_\nu\Phi^0$, and
\begin{equation}
\bar h^\AB \equiv g^\mn\partial_\mu\Phi^A\partial_\nu\Phi^B-\eta^\AB.
\end{equation}
Internal indices (given by capital Roman letters) are raised and lowered with the Minkowski metric. The field equations are\footnote{Note the sign differences between the right-hand side of the Einstein equations and the corresponding equation in \rcite{Chamseddine:2018qym}, which is due to the mostly positive metric convention we employ.}
\begin{align}
G_\mn &= \frac{1}{\Mp^2}T_\mn - \frac{2\lambda}{\Mp^2}\partial_\mu\Phi^0\partial_\nu\Phi^0 \nonumber \\
&\hphantom{{}=}+\frac{m^2}{2}\left(\bar h_\AB-\frac12\bar h\eta_{AB}\right)\left(\partial_\mu\Phi^A\partial_\nu\Phi^B-\frac14 \bar h^\AB g_\mn\right), \label{eq:einstein}\\
0 &= \nabla_\mu\left[\frac{2\lambda}{\Mp^2}\partial^\mu\Phi^0\delta^0_A - \frac{m^2}{2}\left(\bar h_{AB}-\frac12\bar h\eta_{AB}\right)\partial^\mu\Phi^B\right], \\
X &= -1. \label{eq:constraint}
\end{align}
The last of these aligns $\dot\Phi^0$ with the lapse of $g_\mn$. An upshot of this construction is that the constrained mode behaves as a pressureless fluid, i.e., this theory provides a (mimetic) dark matter candidate \cite{Chamseddine:2018qym,Chamseddine:2018gqh}.\footnote{One should note that the phenomenology of mimetic dark matter is still in the early stages of development compared to traditional particle dark matter models such as weakly interacting massive particles (WIMPs) or axions, and it is premature to consider mimetic gravity as a serious alternative to those models. For example, since the mimetic dark matter only interacts gravitationally with the Standard Model, we do not expect to have a thermal production mechanism, in contrast to many traditional dark matter scenarios such as WIMPs. Indeed, when the theory is shift-symmetric in $\Phi^0$, the energy density of this component is set entirely by an integration constant and so is determined by initial conditions. It may also be necessary to tune the parameters of the model in order to obtain the right values of the dark matter density over the entire cosmic history, and higher-derivative effective field theory corrections play an important role \cite{Mirzagholi:2014ifa}. We refer the reader to, e.g., \rcite{Mirzagholi:2014ifa} for discussions of the constraints that early-Universe considerations place on the properties and evolution of mimetic dark matter throughout cosmic history.}

We end this section by making a connection with the existing literature on Lorentz-violating massive gravity and demonstrating the absence of certain well-known features of Lorentz-invariant massive gravity, namely the van Dam-Veltman-Zakharov (vDVZ) discontinuity \cite{vanDam:1970vg,Zakharov:1970cc} and the Higuchi bound \cite{Higuchi:1986py}. The vDVZ discontinuity refers to the failure of linearized Lorentz-invariant massive gravity to reduce to general relativity in the massless limit; this requires nonlinear effects in order to restore general relativity in the Newtonian limit \cite{Vainshtein:1972sx,Babichev:2013usa}. The Higuchi bound is a stability bound for massive gravity on de Sitter space, placing a lower bound on the graviton mass, $m^2\geq 2H^2$, with $H$ the Hubble rate. It is well-known that breaking Lorentz invariance changes both of these conclusions dramatically \cite{Dubovsky:2004sg,Blas:2009my}.

At the level of linear perturbations around flat space, the general SO(3)-invariant mass term in unitary gauge ($\Phi^A=x^A$) can be written as \cite{Dubovsky:2004sg}
\begin{equation}
\mathcal{L}_\mathrm{mass} = \frac18\Mp^2\left(m_0^2h_{00}^2+2m_1^2h_{0i}^2-m_2^2h_{ij}^2+m_3^2h_{ii}^2-2m_4^2h_{00}h_{ii}\right).
\end{equation}
The linearized mass term in \cref{eq:action} in unitary gauge is (treating $\lambda$ as first-order)
\begin{equation}
\mathcal{L}_\mathrm{mass} = \frac{m^2\Mp^2}{8}\left(-\frac12h_{00}^2 + 2 h_{0i}^2-h_{ij}^2+\frac12h_{ii}^2-h_{00}h_{ii}\right) + \lambda h_{00}.
\end{equation}
The $\lambda$ equation of motion sets $h_{00}=0$, which we can impose in the action\footnote{This is justified because, on shell, the $h_{00}$ equation of motion simply sets the value of $\lambda$, while $h_{00}$ drops out of the $h_{ij}$ equations of motion. The dynamics are therefore equivalent.} to find
\begin{equation}
m_0^2 = m_4^2=0,\quad m_1^2=m_2^2=2m_3^2=1.
\end{equation}
This allows us to easily make contact with the existing literature on Lorentz-violating massive gravity. The analysis of \rcite{Dubovsky:2004sg} shows that for these $m_i$ parameters, the Newtonian limit is the usual one, while the vDVZ discontinuity is absent. The analogue of the Higuchi bound in Lorentz-violating massive gravity was derived in \rcite{Blas:2009my}, and for our values of the $m_i$ parameters, it reduces simply to $H^2>0$, which is trivially satisfied.

\section{Flat-space perturbations}
\label{sec:flat}

In this section, we briefly review the behavior of perturbations about flat space in mimetic massive gravity, as discussed in \rcite{Chamseddine:2018qym,Chamseddine:2018gqh}. This will place stability conditions on the theory which will be relevant when we move to cosmological solutions.

The equations of motion \eqref{eq:einstein}--\eqref{eq:constraint} in vacuum are solved by\footnote{This is the only solution that is manifestly invariant under rotations, i.e., with $g_\mn=\diag(-1,1,1,1)$ and $\Phi^A=\left\{\varphi(t),\beta x^i\right\}$. \textit{A priori} it may be possible to have flat solutions with inhomogeneous St\"uckelbergs $\Phi^A$, or equivalently solutions with $\Phi^A=x^A$ and $g_\mn=\eta_\mn$ with $\eta_\mn$ written in a nonstandard coordinate system, but we do not consider these here.}
\begin{equation}
g_\mn = \eta_\mn,\qquad \Phi^A = x^A,\qquad \lambda=0.
\end{equation}
We expand the action \eqref{eq:action} to quadratic order around the Minkowski solution, focusing on scalar modes,
\begin{align}
g_{00} &= -(1+2\phi), \\
g_{0i} &= \partial_iB, \\
g_{ij} &= (1-2\psi)\delta_{ij} + 2\partial_i\partial_j E, \\
\Phi^A &= x^A + \left\{\pi^0,\partial^i\pi\right\}, \\
\lambda &= \delta\lambda.
\end{align}
Three of these fields---$\phi$, $B$, and $\delta\lambda$---are auxiliary, as they appear without time derivatives in the action, and so can be integrated out using their equations of motion. Note that the auxiliary structure is precisely the same as in general relativity, since the mass term and Lagrange multiplier do not introduce any derivatives of the metric.

We can use diffeomorphism invariance to remove a further two modes. When gauge fixing at the level of the action, one must take care to only eliminate variables whose equations of motion are contained in the equations of motion of the remaining variables, otherwise we will lose information after picking a gauge. Following the procedure of \rcite{Lagos:2013aua,Motohashi:2016prk}, we see that we can safely take $\pi^0$ and one of $(E,\pi)$ to vanish. Picking unitary gauge, $\pi^0=\pi=0$, we obtain the flat-space quadratic action (in Fourier space),
\begin{equation}
\delta_2S = \int\dd t\Mp^2\left(-\dot{\vec{\mathcal{X}}}^\mathrm{T}\mathbb{K}\dot{\vec{\mathcal{X}}} + \vec{\mathcal{X}}^\mathrm{T}\left(k^2\mathbb{G}+m^2\mathbb{M}\right)\vec{\mathcal{X}}\right),
\end{equation}
where $\vec{\mathcal{X}}\equiv(\psi,k^2E)$ and the matrices $\mathbb{K}$, $\mathbb{G}$, and $\mathbb M$ are given by
\begin{align}
\mathbb{K} &= \begin{pmatrix}
3+\frac{4k^2}{m^2} & 1 \\
1 & 0
\end{pmatrix}, \\
\mathbb{G} &= \begin{pmatrix}
1 & 0 \\
0 & 0
\end{pmatrix}, \\
\mathbb{M} &= \frac14\begin{pmatrix}
3 & 1 \\
1 & -1
\end{pmatrix}.
\end{align}

As described in \rcite{Chamseddine:2018gqh}, this system can be diagonalized by replacing $\psi$ with the Lagrange multiplier $\delta\lambda$, which we had previously integrated out using
\begin{equation}
\delta\lambda = \frac{\Mp^2}{4}\left[(4k^2+3m^2)\psi+k^2m^2E\right],
\end{equation}
to find
\begin{align}
\delta_2 S &= \int\dd t\frac{1}{4k^2+3m^2}\bigg[k^4m^2\Mp^2\left(\dot E^2-(k^2+m^2)E^2\right)  \nonumber \\
&\hphantom{{}=\int\dd t\frac{1}{4k^2+3m^2}\bigg[}-\frac{1}{\Mp^2}\left(\frac{16}{m^2}\dot{\delta\lambda}^2-4\delta\lambda^2\right)\bigg]. \label{eq:rawquadaction}
\end{align}
If we take $m^2>0$, we can canonically normalize,
\begin{align}
\delta\lambda_\mathrm{c} &\equiv \frac{4}{m \Mp \sqrt{2k^2+\frac32m^2}}\delta\lambda, \label{eq:cannormdl}\\
E_\mathrm{c} &\equiv \frac{m\Mp k^2}{\sqrt{2k^2+\frac32m^2}}E, \label{eq:cannorme}
\end{align}
to obtain the final action,
\begin{equation}
\delta_2 S = \int\dd t\left[\frac12\dot E_\mathrm{c}^2 - \frac12(k^2+m^2)E_\mathrm{c}^2 - \frac12\dot{\delta\lambda}_\mathrm{c}^2 + \frac18m^2\delta\lambda_\mathrm{c}^2\right]. \label{eq:flatspaceaction}
\end{equation}

The only dynamical degree of freedom here is $E_\mathrm{c}$, which is healthy and has mass $m$. The field $\delta\lambda_\mathrm{c}$ has the wrong sign on both its kinetic and mass terms, but does not propagate due to the absence of a gradient term; its equation of motion,
\begin{equation}
\ddot{\delta\lambda}_\mathrm{c} + \frac{m^2}{4}\delta\lambda_\mathrm{c} = 0,
\end{equation}
leads to a dispersion relation $\omega^2 = m^2/4$ and is solved simply by \cite{Chamseddine:2018gqh}
\begin{equation}
\delta\lambda_\mathrm{c} = C(\vec{x})\sin\left(\frac{mt}{2}\right) + D(\vec{x})\cos\left(\frac{mt}{2}\right),
\end{equation}
where $C$ and $D$ are space-dependent constants of integration. The authors of \rcite{Chamseddine:2018gqh} identify this mode with the mimetic dark matter.\footnote{See \rcite{Chamseddine:2018gqh} for an argument for why this mode is not a ghost, despite having an overall wrong-sign action. In principle, one might worry that when quantizing or considering nonlinearities, a coupling will be induced between $\delta\lambda_\mathrm{c}$ and other fields which will lead to an Ostrogradski instability. On the other hand, due to the lack of a gradient term this mode is not a propagating degree of freedom in the usual sense. We will remain agnostic about this question and limit ourselves to considerations of classical, linear stability, which this system clearly satisfies for $m^2>0$. See, e.g., \rcite{Dubovsky:2005xd,Endlich:2010hf} for detailed discussions of classical and quantum properties of modes lacking a gradient term.}

When we discuss cosmology in the next section, we will find ourselves tempted by the possibility of taking $m^2<0$. \textit{A priori} this is merely a parameter choice, but the flat-space analysis shows why this would be a poor decision. By looking at the action \eqref{eq:rawquadaction}, we see that, for negative $m^2$, the overall sign in front of the action flips depending on whether $k^2 > 3|m^2|/4$ or $k^2 < 3|m^2|/4$, a sign of pathological behavior. In particular, for scales $k^2 > 3|m^2|/4$, upon canonically normalizing we find the action \eqref{eq:flatspaceaction} with an overall minus sign, so that the dynamical mode $E_\mathrm{c}$ is a ghost.

%%%%%%%%%%%%%%%%%%%%%%%%%%%%%%%%%%%%%%%%%%%%%%%%%%%%%%%%%%%%
\section{Cosmological solutions}
\label{sec:frw}

In this section we investigate Friedmann-Lema\^{i}tre-Robertson-Walker (FLRW) cosmological solutions of mimetic massive gravity. Consider the homogeneous and isotropic ansatz
\begin{align}
g_\mn &= \diag(-1,a(t)^2\delta_\Ij), \\
\Phi^A &= \left\{\varphi(t),\beta x^i\right\}.
\end{align}
In principle one could allow $\beta$ to depend on time, but this breaks homogeneity and isotropy as it induces $\vec{x}$-dependent terms in the stress-energy tensor of the St\"uckelberg fields. Note that on-shell, the Lagrange multiplier enforces $\varphi=t$ (up to a constant). We will include a general matter sector with density $\rho$ and pressure $p$. We will find this sector needs to contain a cosmological constant, much like in general relativity, but does not need to include dark matter, as this role can be played by the mimetic dark matter (which is an exactly pressureless perfect fluid).

The Einstein and scalar equations of motion are
\begin{align}
3H^2 &= \frac{\rho}{\Mp^2} - \frac{2\lambda}{\Mp^2} - \frac{3m^2}{16}\left(\frac{\beta^4}{a^4}-6\frac{\beta^2}{a^2}+5\right), \label{eq:fried1} \\
2\dot H + 3H^2 &= -\frac{p}{\Mp^2} - \frac{m^2}{16}\left(3-\frac{\beta^4}{a^4}-2\frac{\beta^2}{a^2}\right),\\
0&=\frac{\dd}{\dd t}\left\{a^3\left[\frac{3m^2}{4}\left(1-\frac{\beta^2}{a^2}\right)+\frac{2\lambda}{\Mp^2}\right]\right\} \label{eq:phieom}.
\end{align}
We can solve for $\lambda$ by integrating the $\Phi^0$ equation of motion \eqref{eq:phieom}, finding
\begin{equation}
-\frac{2\lambda}{\Mp^2} = \frac{\mathcal{C}}{a^3} + \frac{3m^2}{4}\left(1-\frac{\beta^2}{a^2}\right), \label{eq:frw-lagrange}
\end{equation}
where $\mathcal{C}$ is an integration constant. Plugging this into the Friedmann equation \eqref{eq:fried1}, we obtain
\begin{equation}
3H^2 = \frac{\rho}{\Mp^2} + \frac{\mathcal{C}}{a^3} - \frac{3m^2}{16}\left(1-\frac{\beta^2}{a^2}\right)^2.
\end{equation}
Note that the contribution from $\lambda$ exactly cancels out that from the last term of the Einstein equation \eqref{eq:einstein}, so the very simple form for $\rho_\varphi \equiv -3m^2\Mp^2(1-\beta^2/a^2)^2/16$ is entirely due to the term proportional to $g_\mn$ in the stress tensor. The integration constant provides a dust-like contribution to the Friedmann equation, which is to be expected as this is a theory of mimetic dark matter.

We can get a better sense of the physical picture by expanding out the Friedmann equation and absorbing the mimetic dark matter $\mathcal{C}$ into $\rho$, finding
\begin{equation}
3H^2 = \frac{\rho}{\Mp^2} - \frac{3m^2}{16}\left(\frac{\beta^4}{a^4}-\frac{2\beta^2}{a^2}+1\right). \label{eq:friedmann}
\end{equation}
For $m^2>0$ ($m^2<0$), we see that the mass term generates an effective negative (positive) cosmological constant, an effective negative (positive) curvature, and an effective radiation component with negative (positive) energy density. Note that these add on to any cosmological constant, radiation, and curvature already present cosmologically; for example, while we have assumed a flat cosmology as our ansatz, observational bounds on spatial curvature will constrain the sum of any pre-existing curvature and the curvature-like term generated by the graviton mass.

Note that for $m^2<0$ we have late-time acceleration, with $\Lambda_\eff=3|m^2|/16$. However, as discussed in the previous section, we need $m^2>0$ in order to avoid a ghost around flat space. This is reminiscent of the situation in the Dvali-Gabadadze-Porrati (DGP) model \cite{Dvali:2000hr}, where one branch of solutions has self-accelerating cosmological expansion \cite{Deffayet:2000uy,Deffayet:2001pu} but is plagued by a ghost \cite{Gorbunov:2005zk,Charmousis:2006pn}, while the other branch is healthy but cannot account for cosmic acceleration.

Let us assume that the energy density $\rho$ in \cref{eq:friedmann} contains dust (including the mimetic dark matter), radiation, and dark energy components. Then, in terms of the density parameters,
\begin{equation}
\Omega_{i,0} = \frac{\rho_{i,0}}{3\Mp^2H_0^2},
\end{equation}
the components of the Friedmann equation which are modified by mimetic massive gravity are
\begin{align}
\Omega_{\Lambda,0} &=\bar{\Omega}_{\Lambda,0} - \frac{m^2}{16H_0^2} \\
\Omega_{K,0}&=\frac{m^2}{8H_0^2}\beta^2, \\
\Omega_{\mathrm{r},0}&=\bar{\Omega}_{\mathrm{r},0} - \frac{m^2}{16H_0^2}\beta^4,
\end{align}
where $\bar{\Omega}_{\Lambda,0}$ and $\bar{\Omega}_{\mathrm{r},0}$ are the densities associated to dark energy and Standard Model radiation. Using observational bounds on the curvature and radiation densities, we can place constraints on the model parameters $m^2$ and $\beta$. We will not consider any bounds coming from the presence of the effective cosmological constant, even though it contributes a negative and potentially large (if $m^2\gg H_0$) amount to $\Omega_{\Lambda,0}$. Particle physics also predicts a large (and potentially negative) vacuum energy, and since we are not worrying about that, it seems inconsistent to worry about the contribution from mimetic massive gravity. One might expect that whatever solves the former problem will also solve the latter.\footnote{See \rcite{Khoury:2018vdv} for a proposed solution to the cosmological constant problem in the context of Lorentz-violating massive gravity, which is closely related to mimetic massive gravity.}

We will use observational constraints on $\Omega_{K,0}$ and $\Omega_{\mathrm{r},0}$ to bound our two free parameters, $m^2$ and $\beta$. \emph{Planck} 2018 constrains $\Omega_{K,0}=0.0007\pm0.0019$, which we parametrize as $|\Omega_{K,0}|<\delta_K$, with $\delta_K\sim 0.003$ \cite{Aghanim:2018eyx}. We will take this to be a constraint on the contribution from mimetic massive gravity alone,
\begin{equation}
\frac{m^2}{8H_0^2}\beta^2 < \delta_K. \label{eq:curv-constraint}
\end{equation}
We remind the reader that what we are really bounding is the sum of the mimetic massive gravity contribution and any ``bare" curvature, but unless there is significant tuning between these two, we can simply take this as a constraint on the mimetic massive gravity piece alone.

To bound the mimetic contribution to the radiation density, we will use constraints from big bang nucleosynthesis (BBN). At the time of BBN, radiation dominates. The exact value of the Hubble rate at the time of nucleosynthesis, which depends on the radiation density, determines the freeze-out abundance of neutrons and therefore the primordial abundance of helium-4, which is subject to tight observational bounds. The constraints are conveniently phrased in terms of the ``speed-up factor" $\zeta\equiv H/\bar H$, where $H$ and $\bar H$ are the Hubble rate and its expected value, respectively, at the time of BBN. The difference between the observed and predicted helium-4 abundance, $|\Delta Y_P|$, is related to the speed-up factor by \cite{Chen:2000xxa}
\begin{equation}
\Delta Y_P = 0.08(\zeta^2-1).\label{eq:DYP}
\end{equation}
Current observational bounds imply \cite{Cyburt:2015mya}
\begin{equation}
|\Delta Y_P|\lesssim 0.01. \label{eq:BBNconst}
\end{equation}
Comparing the Friedmann equation \eqref{eq:friedmann} with and without the mimetic radiation contribution, and focusing on radiation domination, we find
\begin{equation}
\zeta^2 - 1 = -\frac{m^2\beta^4}{16\bar{\Omega}_{\mathrm{r},0}H_0^2},
\end{equation}
where the value for the present-day radiation density associated to photons and neutrinos, $\bar{\Omega}_{\mathrm{r},0}\sim10^{-4}$, is determined entirely by the CMB temperature and the effective number of neutrino species and is therefore not dependent on our modification of gravity.\footnote{See \rcite{2009ApJ...707..916F} for a measurement of the CMB temperature.} Combining this with \cref{eq:BBNconst} we arrive at the constraint
\begin{equation}
\frac{m^2}{16H_0^2}\beta^4 < \delta_\mathrm{r}, \label{eq:rad-constraint}
\end{equation}
where
\begin{equation}
\delta_\mathrm{r} \equiv \frac{\operatorname{max}(|\Delta Y_P|) \bar{\Omega}_{\mathrm{r},0}}{0.08} \approx \mathcal{O}(10^{-5}).
\end{equation}

\begin{figure}
  \includegraphics[width=0.45\textwidth]{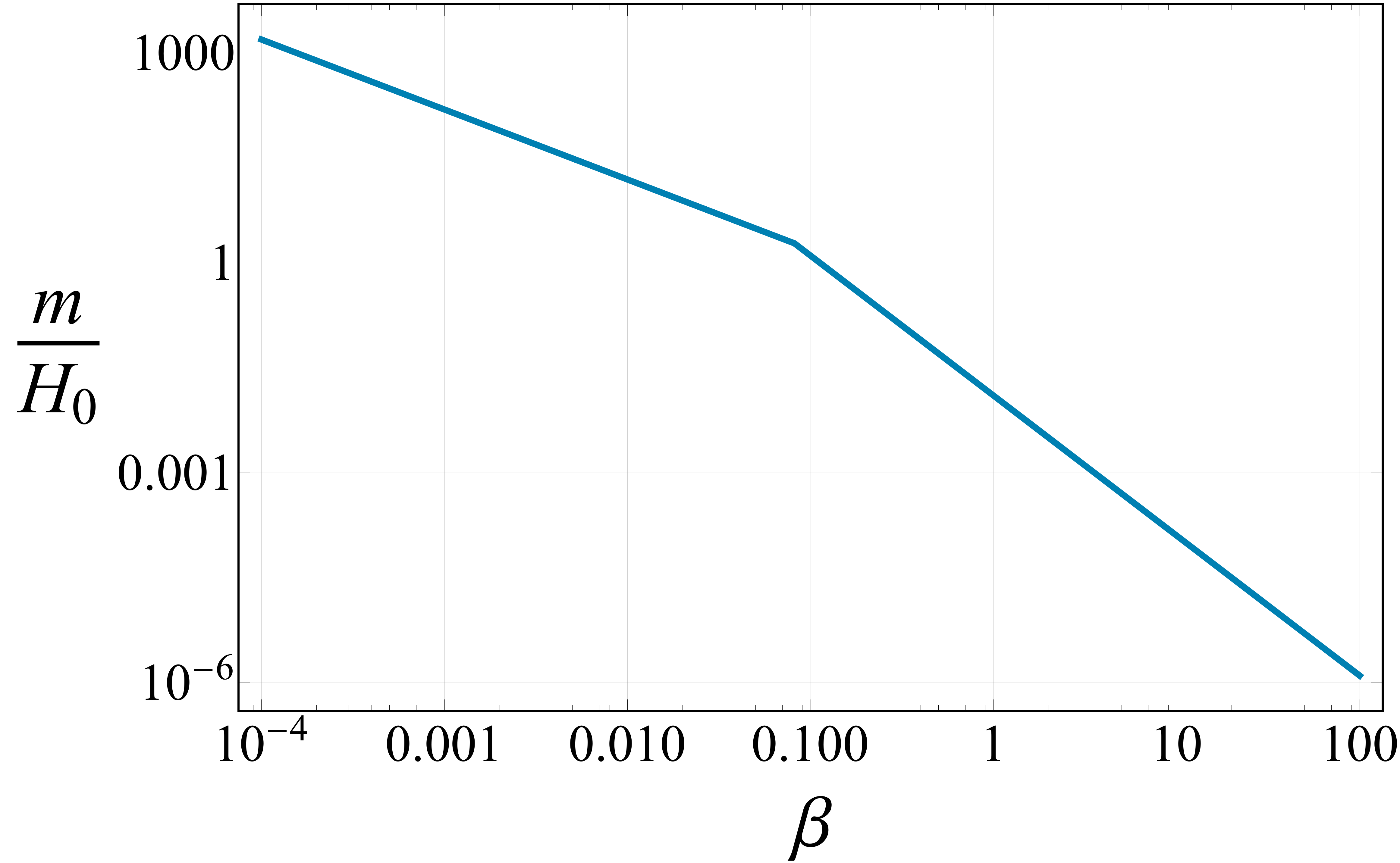}
\caption{Upper limits on $m/H_0$ and $\beta$ for $(\delta_K,\delta_\mathrm{r})=(0.003,10^{-5})$. }
\label{fig:params}
\end{figure}

We can rewrite our constraints \eqref{eq:curv-constraint} and \eqref{eq:rad-constraint} as inequalities for $m/H_0$ and $\beta$ alone in two different r\'egimes,
\begin{equation} \label{eq:mconst-back}
\frac{m}{H_0} < \begin{cases}
\frac{\sqrt{8\delta_K}}{\beta},& \beta < \sqrt{\frac{2\delta_\mathrm{r}}{\delta_K}} \\
\frac{4\sqrt{\delta_\mathrm{r}}}{\beta^2},& \beta > \sqrt{\frac{2\delta_\mathrm{r}}{\delta_K}}.
\end{cases}
\end{equation}
These are plotted in \cref{fig:params}.

Finally, we note that the strong-coupling scale for this theory is of order $\Lambda_2=\sqrt{m\Mp}$ \cite{Chamseddine:2018gqh}. If $m$ is of order the present-day Hubble scale, $m\sim10^{-33}$~eV, then the strong coupling scale is $\Lambda_2\sim$~meV, i.e., the theory breaks down slightly below the millimeter scale. As we see from \cref{eq:mconst-back}, for sufficiently small $\beta$, $m$ could potentially be much larger than $H_0$, leading to a correspondingly larger strong-coupling scale.

\section{Cosmological perturbations}

As we have seen, at the background level, cosmological solutions in mimetic massive gravity do not differ appreciably from $\Lambda$CDM. We therefore proceed to study cosmological perturbations around the FLRW background. This will tell us how cosmological large-scale structure (LSS) evolves in this theory in comparison to $\Lambda$CDM. Since mimetic massive gravity differs from general relativity, we would expect modifications to the gravitational Poisson equation and the slip relation, which could in principle allow for observational tests of this alternative model against $\Lambda$CDM and distinguish the two using the current and future LSS surveys. However, as we will see, stability of cosmological perturbations and the bounds \eqref{eq:mconst-back} place strong constraints on the model which suggest that this theory should be observationally indistinguishable from GR in the linear r\'egime.

\subsection{Stability bound}

We begin by studying the stability of cosmological perturbations using the second-order action formalism. Since, as discussed in \cref{sec:frw}, this theory does not possess ghost-free self-accelerating solutions, we include a cosmological constant, although it will not affect any of the results in this section. Since the theory already contains a pressureless fluid, namely the mimetic dark matter, we need not introduce an additional matter field. Our analysis is therefore valid for all times after matter-radiation equality.

We define the linearized metric, St\"uckelberg fields, and Lagrange multiplier as
\begin{align}
\dd s^2 &= -(1+2\phi)\dd t^2 + 2a\partial_iB\dd t\dd x^i  \nonumber \\
&\hphantom{{}=}+ a^2\left[(1-2\psi)\delta_\Ij+2\partial_i\partial_jE\right]\dd x^i \dd x^j, \\
\Phi^0 &= t + \pi^0, \\
\Phi^i &= \beta\left(x^i + \partial^i\pi\right), \\
\lambda &= \bar\lambda + \delta\lambda,
\end{align}
where we are restricting ourselves to scalar perturbations, and $\bar\lambda$ is the background value given in \cref{eq:frw-lagrange}. The calculation of the quadratic action proceeds analogously to the flat-space case discussed in \cref{sec:flat}. Expanding the action \eqref{eq:action} (with a cosmological constant) to quadratic order in perturbations, we find that the variables $\phi$, $B$, and $\delta\lambda$ are auxiliary---that is, they appear without time derivatives---and can therefore be integrated out using their equations of motion. To safely fix a gauge at the level of the action, we again follow the procedure of \rcite{Lagos:2013aua,Motohashi:2016prk}, finding that we can eliminate one each of ($\psi$,$\pi^0$) and ($E$,$\pi$). We will choose to work in unitary gauge, $\pi^0=\pi=0$, so that $\Phi^A=(t,\beta x^i)$ is unperturbed. The final action, in Fourier space and after integrations by parts, is
\begin{equation}
\delta_2S = \int\dd t\Mp^2a^3\left(-\dot{\vec{\mathcal{X}}}^\mathrm{T}\mathbb{K}\dot{\vec{\mathcal{X}}} + \vec{\mathcal{X}}^\mathrm{T}\left(\frac{k^2}{a^2}\mathbb{G}+m^2\mathbb{M}\right)\vec{\mathcal{X}}\right), \label{eq:pertaction}
\end{equation}
where $\vec{\mathcal{X}}\equiv(\psi,k^2E)$ and the matrices $\mathbb{K}$, $\mathbb{G}$, and $\mathbb M$ are given by
\begin{align}
\mathbb{K} &= \begin{pmatrix}
3-\frac{8a^2}{\beta^2-3a^2}\frac{k^2}{m^2\beta^2} & 1 \\
1 & 0
\end{pmatrix} \\
\mathbb{G} &= \begin{pmatrix}
1 & 0 \\
0 & 0
\end{pmatrix} \\
\mathbb{M} &= \frac18\frac{\beta^2}{a^2}\left(1+\frac{\beta^2}{a^2}\right)\begin{pmatrix}
3 & 1 \\
1 & -1
\end{pmatrix}
\end{align}

Since we are interested in the implications of mimetic massive gravity for the growth and properties of large-scale structure in the late Universe, let us focus on subhorizon scales (i.e., $k^2 \gg a^2H^2$) and assume the quasi-static (QS) approximation. In order to use this approximation, we first need to ensure that fluctuations in this r\'egime are stable. Ignoring time variation in $a(t)$, which will be subdominant in the limit $k^2 \gg a^2H^2$, and assuming solutions of the form $\vec{\mathcal{X}}=\vec{\mathcal{X}}_0e^{i\omega t}$, the equations of motion following from the action \eqref{eq:pertaction} are
\begin{equation}
\left(-\omega^2\mathbb{K}+\frac{k^2}{a^2}\mathbb{G}+m^2\mathbb{M}\right)\vec{\mathcal{X}}=0.
\end{equation}
We can then derive stability conditions from the dispersion relations, obtained by solving
\begin{widetext}
\begin{align} \label{eq:disp}
0 &= \det\left(-\omega^2\mathbb{K}+\frac{k^2}{a^2}\mathbb{G}+m^2\mathbb{M}\right) \nonumber \\
&= \frac{\omega^4}{a^2+\beta^2} - \frac{\omega^2k^2}{a^2(3a^2-\beta^2)} - \frac{5\omega^2m^2\beta^2}{8a^4} + \frac{k^2m^2\beta^2}{8a^6} + \frac{m^4\beta^4(a^2+\beta^2)}{16a^8}
\end{align}
\end{widetext}
for $\omega^2$.

The dispersion relations arising from \cref{eq:disp} are complicated, but simplify significantly in the limit $k\gg aH$ when we take into account the constraints \eqref{eq:mconst-back} on $m/H_0$, which we obtained by requiring that the radiation and curvature densities generated by the mass term not exceed observational bounds. Consider replacing $m$ and $\beta$ in \cref{eq:disp} with the following two parameters,\footnote{To do this replacement, first replace $m\to\sqrt{\epsilon_1}\beta/k$, and then replace any remaining factors of $\beta$ with $\beta\to\sqrt{\epsilon_2/\epsilon_1}a$.}
\begin{equation}
\epsilon_1\equiv\left(\frac{m\beta}{k}\right)^2,\qquad\epsilon_2\equiv\left(\frac{m\beta^2}{ka}\right)^2.
\end{equation}
We proceed to show that the bounds \eqref{eq:mconst-back} imply that each of these is much smaller than unity on subhorizon scales for all times after matter-radiation equality.

For both $\epsilon_1$ and $\epsilon_2$ we can put upper bounds on the numerators and lower bounds on the denominators. Let us start with the numerators. For $\epsilon_1$, multiply each side of \cref{eq:mconst-back} by $\beta$. We see there is a strict upper bound on the combination $m\beta$,
\begin{equation}
m\beta\leq\sqrt{8\delta_K}H_0\approx0.15H_0 \label{eq:mbconst}
\end{equation}
where we have taken $\delta_K\sim0.003$ as a representative value. We can similarly find a bound on the numerator of $\epsilon_2$ by multiplying both sides of \cref{eq:mconst-back} by $\beta^2$, finding
\begin{equation}
m\beta^2\leq4\sqrt{\delta_r}H_0\approx10^{-2}H_0
\end{equation}
for $\delta_r\sim10^{-5}$.

Now we move on to the denominators. The subhorizon limit is given by $k\gg aH$. For the sake of argument let us be conservative and assume that $k$ is only slightly subhorizon, $k/a\approx \mathcal{O}(1) H$.\footnote{Of course, the deeper in the subhorizon r\'egime $k$ is, the smaller $\epsilon_1$ and $\epsilon_2$ become.} At any given time from matter-radiation equality to the present, where we can trust our analysis, the Hubble rate $H$ is related to its present-day value $H_0$ by $H=H_0\sqrt{\Omega_{\Lambda,0}+\Omega_{\mathrm{m},0}a^{-3}}$. Putting this together with the bounds we have derived on $m\beta$ and $m\beta^2$, we find
\begin{align}
\epsilon_1 &\lesssim \frac{0.02}{\Omega_{\Lambda,0}a^2+\Omega_{\mathrm{m},0}a^{-1}} \ll 1, \\
\epsilon_2 &\lesssim \frac{10^{-4}}{\Omega_{\Lambda,0}a^4+\Omega_{\mathrm{m},0}a} \ll 1\text{ for }z \lesssim3000.
\end{align}
Note that while the upper bound on $\epsilon_1$ is always much smaller than unity for $0<a\leq1$, the upper bound on $\epsilon_2$ in fact grows as $a^{-1}$ at early times. However, it grows slowly and has a factor of $10^{-4}$ to compete with, so that $\max(\epsilon_2)$ does not reach unity until $z\sim3000$, right around matter-radiation equality. Therefore in principle there might be a handful of modes---right around the horizon scale and at the earliest moments of matter domination---for which terms going as $\epsilon_2$ affect the subhorizon dispersion relation, \emph{if} $m\beta^2$ takes the largest value allowed by the constraints. We will continue to take $\epsilon_2\ll1$, with the understanding that if this particular situation is realized, then at those very early times we are only considering modes with $k\gtrsim10aH$, for which $\epsilon_2$ is certainly smaller than unity.

Dropping terms subdominant in $\epsilon_1$ and $\epsilon_2$, the dispersion relation \eqref{eq:disp} becomes
\begin{equation}
0 \approx \frac{\omega^4}{a^2+\beta^2} - \frac{\omega^2k^2}{a^2(3a^2-\beta^2)} + \frac{k^2m^2\beta^2}{8a^6}.
\end{equation}
Solving for $\omega^2$, and again dropping terms subleading in $\epsilon_1=(m\beta/k)^2$, we find the dispersion relations for our two modes,
\begin{align}
\omega^2 &\approx \frac{k^2}{a^2}\frac{a^2+\beta^2}{3a^2-\beta^2}, \\
\omega^2 &\approx \frac{m^2\beta^2}{8a^2}\left(3-\frac{\beta^2}{a^2}\right).
\end{align}
Each of these implies the same stability condition,
\begin{equation}
\frac{\beta^2}{a^2}<3. \label{eq:stab}
\end{equation}
This tells us that no matter what the value of $\beta$ is, our cosmological solutions are \emph{unstable} at sufficiently high redshifts,
\begin{equation}
z>\sqrt{3}\beta^{-1}-1. 
\end{equation}
This early time instability can however be safely pushed back to unobservably early times by taking the parameter $\beta$ to be sufficiently small.\footnote{This is similar to massive bimetric gravity, which possesses an early-time instability that can be rendered safe in the limit where the ratio of the two Planck masses becomes small~\cite{Akrami:2015qga}.} Because we are assuming matter and dark energy domination, we can trust our stability condition as far back as matter-radiation equality at $z_\mathrm{eq}\approx3400$. Demanding stability from $z_\mathrm{eq}$ onward, we find a constraint on $\beta$,\footnote{It is plausible that the result \eqref{eq:stab} holds, at least on an order-of-magnitude basis, through radiation domination as well (see, again, the example of bigravity \cite{Akrami:2015qga}). In this case, we should demand that the instability be pushed back to before big bang nucleosynthesis, with $z_\mathrm{BBN} \approx 3 \times 10^8$, which would imply a stronger constraint of $\beta \lesssim 10^{-8}$. We do not have much observational handle on the presumably radiation-dominated era before BBN, and therefore should not demand that the instability be absent then; indeed, a mild enough instability might have interesting consequences, such as the formation of primordial black holes.}
\begin{equation}
\beta\lesssim5\times10^{-4}. \label{eq:betaconst}
\end{equation}

\subsection{Cosmological tensor mass}

Another possible cosmological bound on the parameters $m$ and $\beta$ comes from constraints on the graviton mass. The tightest bounds currently come from LIGO, $m_\mathrm{T}\leq7.7\times10^{-23}$~eV \cite{Abbott:2017vtc}.\footnote{See \rcite{deRham:2016nuf} for a helpful summary of bounds on the graviton mass from a variety of experiments and observations.} To compute the mass of tensor fluctuations on a cosmological background, we linearize the Einstein equation \eqref{eq:einstein} around $g_\mn  = \bar{g}_\mn+h_\mn$, with $\bar{g}_\mn=\diag(-1,a^2\delta_\Ij)$, $h_{00}=0$, and $h_\Ij$ transverse and traceless, i.e., $h_{ii}=\partial_ih_\Ij=0$. The Einstein equation is
\begin{equation}
\ddot h_\Ij + 3H\dot h_\Ij -\frac{\nabla^2}{a^2}h_\Ij + m_\mathrm{T}^2h_\Ij=0
\end{equation}
with the tensor mass given by
\begin{equation}
m_\mathrm{T}^2 \equiv \frac{m^2}{2}\frac{\beta^2}{a^2}\left(1+\frac{\beta^2}{a^2}\right)
\end{equation}

The structure of the Einstein equation is such that $m_\mathrm{T}^2/m^2$ has to be a (quadratic) polynomial in $\beta^2/a^2$. What is non-trivial is that the degree-zero term in that polynomial cancels out, i.e., the expression for $m_\mathrm{T}^2/m^2$ starts at order $\beta^2/a^2$. This means that gravitational waves propagating over cosmological distances (at low redshift, i.e., $a\sim\mathcal{O}(1)$) do not depend on $m$ alone; instead they involve the combinations $m\beta$ and $m\beta^2$ which, as we have seen, are strongly constrained by the cosmological background. In particular, recalling that $m^2\beta^2\lesssim10^{-2}H_0^2$ and $m^2\beta^4\lesssim10^{-4}H_0^2$, we see that $m_\mathrm{T}$ at the present era is at most of order $10^{-1}H_0 \sim 10^{-34}$~eV, far below the LIGO bounds. Moreover, our stability condition \eqref{eq:betaconst} has no bearing on $m_\mathrm{T}$. No matter how tiny $\beta$ is, the constraints \eqref{eq:mconst-back} place a constant upper bound on $m\beta$, so that the smaller $\beta$ is, the larger $m$ is allowed to be, leaving $m_\mathrm{T}\approx m\beta/(\sqrt2a)$ fixed. It is interesting to note that, without demanding that this model provide cosmic acceleration, the tensor mass is nevertheless forced to be smaller than the Hubble scale. Finally, we note that around a flat background, the tensor mass is simply $m$, so local tests of gravity might be able to place constraints on $m$ that are not possible with gravitational waves that propagate over cosmological distances.

\subsection{Quasistatic limit}

Finally, let us comment on the testability of mimetic massive gravity using near-future LSS surveys. We will find it convenient to work in Newtonian gauge, $B=E=0$. Linearizing the Einstein equations \eqref{eq:einstein}, and leaving in a generic stress-energy tensor $T_\mn$ for completeness, we obtain
\begin{widetext}
\begin{align}
6H^2\phi - \frac{2}{a^2}\partial^i\partial_i\psi + 6H\dot{\psi} &= \frac{1}{\Mp^2}\delta T^0_{\,0} + 2\frac{\delta\lambda}{\Mp^2}- \frac{m^2}{4}\frac{\beta^2}{a^2}\left(3-\frac{\beta^2}{a^2}\right)\left(3\psi + \partial^i\partial_i\pi\right),\label{eq:0-0}\\
-2\partial_i\left(\dot\psi+H\phi \right) &= \frac{1}{\Mp^2}\delta T^0_{\,i} + 2\frac{\bar\lambda}{\Mp^2}\partial_i\pi^0 + \frac{m^2}{4}\left(3-\frac{\beta^2}{a^2}\right)\left(\partial_i\pi^0 -\beta^2\partial_i\dot\pi\right),\label{eq:0-i}\\
6\left[\ddot \psi+3H\dot\psi+H\dot\phi+(3H^2+2\dot H)\phi\right] + \frac{2}{a^2}\partial^i\partial_i\left(\phi-\psi \right) &= \frac{1}{\Mp^2}\delta T^i_{\,i} - \frac{m^2}{4}\frac{\beta^2}{a^2}\left(1+\frac{\beta^2}{a^2}\right)\left(3\psi + \partial^i\partial_i\pi \right),\label{eq:trace}\\
\frac{1}{a^2}\partial^i\partial_j\left( \psi - \phi\right)&= \frac{1}{\Mp^2}\delta T^i_{\,j} + \frac{m^2}{2}\frac{\beta^2}{a^2}\left(1+\frac{\beta^2}{a^2}\right)\partial^i\partial_j\pi,\quad i\neq j.\label{eq:i-j}
\end{align}
\end{widetext}
Moving to Fourier space, specializing to a pressureless fluid without anisotropic stress, and taking the quasistatic limit, $\ddot X\sim H\dot X\sim H^2X\ll k^2X$ for any perturbation $X$, \cref{eq:0-0,eq:i-j,eq:trace} become
\begin{align}
\frac{2k^2}{a^2}\psi &= \frac{1}{\Mp^2}\left(2\delta\lambda-\bar\rho\delta\right) \nonumber \\
&\hphantom{{}=}- \frac{m^2}{4}\frac{\beta^2}{a^2}\left(3-\frac{\beta^2}{a^2}\right)\left(3\psi-k^2\pi\right),\label{eq:0-0-QS}\\
\frac{2k^2}{a^2}\left(\phi-\psi \right) &= \frac{m^2}{4}\frac{\beta^2}{a^2}\left(1+\frac{\beta^2}{a^2}\right)\left(3\psi -k^2\pi \right), \label{eq:trace-QS}\\
\frac{1}{a^2}\left( \phi - \psi\right)&= -\frac{m^2}{2}\frac{\beta^2}{a^2}\left(1+\frac{\beta^2}{a^2}\right)\pi, \label{eq:i-j-QS}
\end{align}
where $\bar\rho$ and $\delta$ are the background density and overdensity of the dust component. Note that these are degenerate with the mimetic dark matter, as expected.

Combining these equations, we obtain the modified Poisson equation and the slip relation,
\begin{align}
-k^2\psi &= 4\pi G \mu(a,k) a^2(\delta\rho-2\delta\lambda),\label{eq:poisson-QS}\\
\psi &= \eta(a,k)\phi,\label{eq:slip-QS}
\end{align}
where the modified-gravity parameters $\mu$ and $\eta$ are given by
\begin{align}
\mu(a,k) &= \frac{1}{1+\frac{1}{2}\frac{m^2\beta^2}{k^2}\left(3-\frac{\beta^2}{a^2}\right)},\label{eq:mu}\\
\eta(a,k) &= \frac{1}{1+\frac{1}{2}\frac{m^2\beta^2}{k^2}\left(1+\frac{\beta^2}{a^2}\right)}.\label{eq:eta}
\end{align}
These parametrize observable deviations from general relativity, in which $\mu=\eta=1$.

The constraints we have already derived on $m$ and $\beta$ preclude $\mu$ and $\eta$ from deviating from unity at a level accessible to near-future observations. The stability constraint \eqref{eq:stab} requires the terms in parentheses to be $\mathcal{O}(1)$, while the background constraint \eqref{eq:mbconst} sets $m^2\beta^2\lesssim0.02H_0^2$, so that
\begin{equation}
\mu-1\sim\eta-1\sim\mathcal{O}\left(\frac{m^2\beta^2}{k^2}\right)\lesssim10^{-2}\left(\frac{H_0}{k}\right)^2.
\end{equation}
It is therefore highly unlikely that cosmological observations will be able to test this model against $\Lambda$CDM in the linear and subhorizon r\'egime.

\section{Conclusions}

In this Letter we have studied the first cosmological implications of the recently-proposed theory of mimetic massive gravity. We find that the theory is unable to self-accelerate without introducing a ghost. Its effects on Friedmann-Lema\^{i}tre-Robertson-Walker cosmological backgrounds are to introduce effective radiation, curvature, and cosmological constant terms, as well as a dust-like mimetic dark matter component. We place constraints \eqref{eq:mconst-back} on the theory parameters by demanding that the effective radiation and curvature terms be within observational bounds. In the ghost-free region of parameter space, $m^2>0$, the effective cosmological constant is negative-definite, so a separate dark energy sector, which we take to be a positive cosmological constant, is required to explain the late-time acceleration of the Universe.

We further studied the behavior of cosmological perturbations in the subhorizon, quasistatic limit. The model generically suffers from an instability at early times. However, since our analysis only included a pressureless dust component (in addition to a cosmological constant), the calculation can only be trusted as far back as matter-radiation equality. This allowed us to place a further constraint on the theory parameters by insisting that the instability be absent throughout matter domination. With these constraints, the deviations from $\Lambda$CDM in the linear, subhorizon r\'egime are likely too small to be observable.

Not surprisingly, since this is a theory of massive gravity, it predicts massive tensor modes. We have calculated the tensor mass around cosmological backgrounds and found that, taking into account the constraints imposed by the cosmological background, this mass must be at least an order of magnitude below the Hubble scale, far outside the currently-available constraints on the graviton mass. Unlike other theories of massive gravity, in which the graviton mass is comparable to the Hubble scale in order to provide late-time acceleration, this bound on the graviton mass is solely due to the requirement that the effective radiation and curvature terms in the Friedmann equation not be too large.

What are the remaining prospects for cosmological tests of mimetic massive gravity? We emphasize that our analysis does not apply in two important r\'egimes: horizon-size scales and nonlinear scales. One or both of these may possess signatures which could be used to distinguish mimetic massive gravity from $\Lambda$CDM, or otherwise to rule out additional regions of parameter space. One expects that nonlinear scales will require $N$-body simulations, while at horizon-size scales we cannot apply the quasistatic approximation and would need to solve the perturbation equations numerically, as in other theories of modified gravity \cite{Enander:2015vja}. For the latter, we note that the mass scales appearing in the action \eqref{eq:pertaction} for cosmological perturbations are not simply $m$, which can be arbitrarily large (in the limit of small $\beta$), but rather $m \beta$ and $m\beta^2$, which we have shown must both be at least an order of magnitude smaller than the Hubble scale. It therefore might be difficult for this theory to produce effects at horizon scales that are larger than cosmic variance. Note that scales $k\sim m\beta$ and $k\sim m\beta^2$ are super-horizon and therefore not observable.

\begin{acknowledgments}
We are grateful to Marco Crisostomi, Sergei Dubovsky, Nick Kaiser, Riccardo Penco, Jeremy Sakstein, and Filippo Vernizzi for helpful discussions. We thank the referee for useful comments on the draft. A.R.S. is supported by DOE HEP grants DOE DE-FG02-04ER41338 and FG02-06ER41449 and by the McWilliams Center for Cosmology, Carnegie Mellon University. V.V. is supported by a de Sitter PhD fellowship of the Netherlands Organization for Scientific Research (NWO). Y.A. is supported by LabEx ENS-ICFP: ANR-10-LABX-0010/ANR-10-IDEX-0001-02 PSL*. Y.A. also acknowledges support from the NWO and the Dutch Ministry of Education, Culture and Science (OCW), and also from the D-ITP consortium, a program of the NWO that is funded by the OCW. 
\end{acknowledgments}

\bibliography{refs}

\begin{thebibliography}{46}%
\makeatletter
\providecommand \@ifxundefined [1]{%
 \@ifx{#1\undefined}
}%
\providecommand \@ifnum [1]{%
 \ifnum #1\expandafter \@firstoftwo
 \else \expandafter \@secondoftwo
 \fi
}%
\providecommand \@ifx [1]{%
 \ifx #1\expandafter \@firstoftwo
 \else \expandafter \@secondoftwo
 \fi
}%
\providecommand \natexlab [1]{#1}%
\providecommand \enquote  [1]{``#1''}%
\providecommand \bibnamefont  [1]{#1}%
\providecommand \bibfnamefont [1]{#1}%
\providecommand \citenamefont [1]{#1}%
\providecommand \href@noop [0]{\@secondoftwo}%
\providecommand \href [0]{\begingroup \@sanitize@url \@href}%
\providecommand \@href[1]{\@@startlink{#1}\@@href}%
\providecommand \@@href[1]{\endgroup#1\@@endlink}%
\providecommand \@sanitize@url [0]{\catcode `\\12\catcode `\$12\catcode
  `\&12\catcode `\#12\catcode `\^12\catcode `\_12\catcode `\%12\relax}%
\providecommand \@@startlink[1]{}%
\providecommand \@@endlink[0]{}%
\providecommand \url  [0]{\begingroup\@sanitize@url \@url }%
\providecommand \@url [1]{\endgroup\@href {#1}{\urlprefix }}%
\providecommand \urlprefix  [0]{URL }%
\providecommand \Eprint [0]{\href }%
\providecommand \doibase [0]{http://dx.doi.org/}%
\providecommand \selectlanguage [0]{\@gobble}%
\providecommand \bibinfo  [0]{\@secondoftwo}%
\providecommand \bibfield  [0]{\@secondoftwo}%
\providecommand \translation [1]{[#1]}%
\providecommand \BibitemOpen [0]{}%
\providecommand \bibitemStop [0]{}%
\providecommand \bibitemNoStop [0]{.\EOS\space}%
\providecommand \EOS [0]{\spacefactor3000\relax}%
\providecommand \BibitemShut  [1]{\csname bibitem#1\endcsname}%
\let\auto@bib@innerbib\@empty
%</preamble>
\bibitem [{\citenamefont {Chamseddine}\ and\ \citenamefont
  {Mukhanov}(2018{\natexlab{a}})}]{Chamseddine:2018qym}%
  \BibitemOpen
  \bibfield  {author} {\bibinfo {author} {\bibfnamefont {A.~H.}\ \bibnamefont
  {Chamseddine}}\ and\ \bibinfo {author} {\bibfnamefont {V.}~\bibnamefont
  {Mukhanov}},\ }\href {\doibase 10.1007/JHEP06(2018)060} {\bibfield  {journal}
  {\bibinfo  {journal} {JHEP}\ }\textbf {\bibinfo {volume} {06}},\ \bibinfo
  {pages} {060} (\bibinfo {year} {2018}{\natexlab{a}})},\ \Eprint
  {http://arxiv.org/abs/1805.06283} {arXiv:1805.06283 [hep-th]} \BibitemShut
  {NoStop}%
%%CITATION = ARXIV:1805.06283;%%
\bibitem [{\citenamefont {Chamseddine}\ and\ \citenamefont
  {Mukhanov}(2018{\natexlab{b}})}]{Chamseddine:2018gqh}%
  \BibitemOpen
  \bibfield  {author} {\bibinfo {author} {\bibfnamefont {A.~H.}\ \bibnamefont
  {Chamseddine}}\ and\ \bibinfo {author} {\bibfnamefont {V.}~\bibnamefont
  {Mukhanov}},\ }\href {\doibase 10.1007/JHEP06(2018)062} {\bibfield  {journal}
  {\bibinfo  {journal} {JHEP}\ }\textbf {\bibinfo {volume} {06}},\ \bibinfo
  {pages} {062} (\bibinfo {year} {2018}{\natexlab{b}})},\ \Eprint
  {http://arxiv.org/abs/1805.06598} {arXiv:1805.06598 [hep-th]} \BibitemShut
  {NoStop}%
%%CITATION = ARXIV:1805.06598;%%
\bibitem [{\citenamefont {Chamseddine}\ and\ \citenamefont
  {Mukhanov}(2013)}]{Chamseddine:2013kea}%
  \BibitemOpen
  \bibfield  {author} {\bibinfo {author} {\bibfnamefont {A.~H.}\ \bibnamefont
  {Chamseddine}}\ and\ \bibinfo {author} {\bibfnamefont {V.}~\bibnamefont
  {Mukhanov}},\ }\href {\doibase 10.1007/JHEP11(2013)135} {\bibfield  {journal}
  {\bibinfo  {journal} {JHEP}\ }\textbf {\bibinfo {volume} {11}},\ \bibinfo
  {pages} {135} (\bibinfo {year} {2013})},\ \Eprint
  {http://arxiv.org/abs/1308.5410} {arXiv:1308.5410 [astro-ph.CO]} \BibitemShut
  {NoStop}%
%%CITATION = ARXIV:1308.5410;%%
\bibitem [{\citenamefont {Fierz}\ and\ \citenamefont
  {Pauli}(1939)}]{Fierz:1939ix}%
  \BibitemOpen
  \bibfield  {author} {\bibinfo {author} {\bibfnamefont {M.}~\bibnamefont
  {Fierz}}\ and\ \bibinfo {author} {\bibfnamefont {W.}~\bibnamefont {Pauli}},\
  }\href {\doibase 10.1098/rspa.1939.0140} {\bibfield  {journal} {\bibinfo
  {journal} {Proc.Roy.Soc.Lond.}\ }\textbf {\bibinfo {volume} {A173}},\
  \bibinfo {pages} {211} (\bibinfo {year} {1939})}\BibitemShut {NoStop}%
%%CITATION = PRSLA,A173,211;%%
\bibitem [{\citenamefont {Isham}\ \emph {et~al.}(1971)\citenamefont {Isham},
  \citenamefont {Salam},\ and\ \citenamefont {Strathdee}}]{Isham:1971gm}%
  \BibitemOpen
  \bibfield  {author} {\bibinfo {author} {\bibfnamefont {C.}~\bibnamefont
  {Isham}}, \bibinfo {author} {\bibfnamefont {A.}~\bibnamefont {Salam}}, \ and\
  \bibinfo {author} {\bibfnamefont {J.}~\bibnamefont {Strathdee}},\ }\href
  {\doibase 10.1103/PhysRevD.3.867} {\bibfield  {journal} {\bibinfo  {journal}
  {Phys.Rev.}\ }\textbf {\bibinfo {volume} {D3}},\ \bibinfo {pages} {867}
  (\bibinfo {year} {1971})}\BibitemShut {NoStop}%
%%CITATION = PHRVA,D3,867;%%
\bibitem [{\citenamefont {Boulware}\ and\ \citenamefont
  {Deser}(1972)}]{Boulware:1973my}%
  \BibitemOpen
  \bibfield  {author} {\bibinfo {author} {\bibfnamefont {D.}~\bibnamefont
  {Boulware}}\ and\ \bibinfo {author} {\bibfnamefont {S.}~\bibnamefont
  {Deser}},\ }\href {\doibase 10.1103/PhysRevD.6.3368} {\bibfield  {journal}
  {\bibinfo  {journal} {Phys.Rev.}\ }\textbf {\bibinfo {volume} {D6}},\
  \bibinfo {pages} {3368} (\bibinfo {year} {1972})}\BibitemShut {NoStop}%
%%CITATION = PHRVA,D6,3368;%%
\bibitem [{\citenamefont {Creminelli}\ \emph {et~al.}(2005)\citenamefont
  {Creminelli}, \citenamefont {Nicolis}, \citenamefont {Papucci},\ and\
  \citenamefont {Trincherini}}]{Creminelli:2005qk}%
  \BibitemOpen
  \bibfield  {author} {\bibinfo {author} {\bibfnamefont {P.}~\bibnamefont
  {Creminelli}}, \bibinfo {author} {\bibfnamefont {A.}~\bibnamefont {Nicolis}},
  \bibinfo {author} {\bibfnamefont {M.}~\bibnamefont {Papucci}}, \ and\
  \bibinfo {author} {\bibfnamefont {E.}~\bibnamefont {Trincherini}},\ }\href
  {\doibase 10.1088/1126-6708/2005/09/003} {\bibfield  {journal} {\bibinfo
  {journal} {JHEP}\ }\textbf {\bibinfo {volume} {0509}},\ \bibinfo {pages}
  {003} (\bibinfo {year} {2005})},\ \Eprint
  {http://arxiv.org/abs/hep-th/0505147} {arXiv:hep-th/0505147 [hep-th]}
  \BibitemShut {NoStop}%
%%CITATION = HEP-TH/0505147;%%
\bibitem [{\citenamefont {de~Rham}\ and\ \citenamefont
  {Gabadadze}(2010)}]{deRham:2010ik}%
  \BibitemOpen
  \bibfield  {author} {\bibinfo {author} {\bibfnamefont {C.}~\bibnamefont
  {de~Rham}}\ and\ \bibinfo {author} {\bibfnamefont {G.}~\bibnamefont
  {Gabadadze}},\ }\href {\doibase 10.1103/PhysRevD.82.044020} {\bibfield
  {journal} {\bibinfo  {journal} {Phys.Rev.}\ }\textbf {\bibinfo {volume}
  {D82}},\ \bibinfo {pages} {044020} (\bibinfo {year} {2010})},\ \Eprint
  {http://arxiv.org/abs/1007.0443} {arXiv:1007.0443 [hep-th]} \BibitemShut
  {NoStop}%
%%CITATION = ARXIV:1007.0443;%%
\bibitem [{\citenamefont {de~Rham}\ \emph {et~al.}(2011)\citenamefont
  {de~Rham}, \citenamefont {Gabadadze},\ and\ \citenamefont
  {Tolley}}]{deRham:2010kj}%
  \BibitemOpen
  \bibfield  {author} {\bibinfo {author} {\bibfnamefont {C.}~\bibnamefont
  {de~Rham}}, \bibinfo {author} {\bibfnamefont {G.}~\bibnamefont {Gabadadze}},
  \ and\ \bibinfo {author} {\bibfnamefont {A.~J.}\ \bibnamefont {Tolley}},\
  }\href {\doibase 10.1103/PhysRevLett.106.231101} {\bibfield  {journal}
  {\bibinfo  {journal} {Phys.Rev.Lett.}\ }\textbf {\bibinfo {volume} {106}},\
  \bibinfo {pages} {231101} (\bibinfo {year} {2011})},\ \Eprint
  {http://arxiv.org/abs/1011.1232} {arXiv:1011.1232 [hep-th]} \BibitemShut
  {NoStop}%
%%CITATION = ARXIV:1011.1232;%%
\bibitem [{\citenamefont {Hassan}\ and\ \citenamefont
  {Rosen}(2012{\natexlab{a}})}]{Hassan:2011hr}%
  \BibitemOpen
  \bibfield  {author} {\bibinfo {author} {\bibfnamefont {S.}~\bibnamefont
  {Hassan}}\ and\ \bibinfo {author} {\bibfnamefont {R.~A.}\ \bibnamefont
  {Rosen}},\ }\href {\doibase 10.1103/PhysRevLett.108.041101} {\bibfield
  {journal} {\bibinfo  {journal} {Phys.Rev.Lett.}\ }\textbf {\bibinfo {volume}
  {108}},\ \bibinfo {pages} {041101} (\bibinfo {year} {2012}{\natexlab{a}})},\
  \Eprint {http://arxiv.org/abs/1106.3344} {arXiv:1106.3344 [hep-th]}
  \BibitemShut {NoStop}%
%%CITATION = ARXIV:1106.3344;%%
\bibitem [{\citenamefont {Hassan}\ \emph
  {et~al.}(2012{\natexlab{a}})\citenamefont {Hassan}, \citenamefont {Rosen},\
  and\ \citenamefont {Schmidt-May}}]{Hassan:2011tf}%
  \BibitemOpen
  \bibfield  {author} {\bibinfo {author} {\bibfnamefont {S.}~\bibnamefont
  {Hassan}}, \bibinfo {author} {\bibfnamefont {R.~A.}\ \bibnamefont {Rosen}}, \
  and\ \bibinfo {author} {\bibfnamefont {A.}~\bibnamefont {Schmidt-May}},\
  }\href {\doibase 10.1007/JHEP02(2012)026} {\bibfield  {journal} {\bibinfo
  {journal} {JHEP}\ }\textbf {\bibinfo {volume} {1202}},\ \bibinfo {pages}
  {026} (\bibinfo {year} {2012}{\natexlab{a}})},\ \Eprint
  {http://arxiv.org/abs/1109.3230} {arXiv:1109.3230 [hep-th]} \BibitemShut
  {NoStop}%
%%CITATION = ARXIV:1109.3230;%%
\bibitem [{\citenamefont {Hassan}\ and\ \citenamefont
  {Rosen}(2012{\natexlab{b}})}]{Hassan:2011ea}%
  \BibitemOpen
  \bibfield  {author} {\bibinfo {author} {\bibfnamefont {S.}~\bibnamefont
  {Hassan}}\ and\ \bibinfo {author} {\bibfnamefont {R.~A.}\ \bibnamefont
  {Rosen}},\ }\href {\doibase 10.1007/JHEP04(2012)123} {\bibfield  {journal}
  {\bibinfo  {journal} {JHEP}\ }\textbf {\bibinfo {volume} {1204}},\ \bibinfo
  {pages} {123} (\bibinfo {year} {2012}{\natexlab{b}})},\ \Eprint
  {http://arxiv.org/abs/1111.2070} {arXiv:1111.2070 [hep-th]} \BibitemShut
  {NoStop}%
%%CITATION = ARXIV:1111.2070;%%
\bibitem [{\citenamefont {Hassan}\ \emph
  {et~al.}(2012{\natexlab{b}})\citenamefont {Hassan}, \citenamefont
  {Schmidt-May},\ and\ \citenamefont {von Strauss}}]{Hassan:2012qv}%
  \BibitemOpen
  \bibfield  {author} {\bibinfo {author} {\bibfnamefont {S.}~\bibnamefont
  {Hassan}}, \bibinfo {author} {\bibfnamefont {A.}~\bibnamefont {Schmidt-May}},
  \ and\ \bibinfo {author} {\bibfnamefont {M.}~\bibnamefont {von Strauss}},\
  }\href {\doibase 10.1016/j.physletb.2012.07.018} {\bibfield  {journal}
  {\bibinfo  {journal} {Phys.Lett.}\ }\textbf {\bibinfo {volume} {B715}},\
  \bibinfo {pages} {335} (\bibinfo {year} {2012}{\natexlab{b}})},\ \Eprint
  {http://arxiv.org/abs/1203.5283} {arXiv:1203.5283 [hep-th]} \BibitemShut
  {NoStop}%
%%CITATION = ARXIV:1203.5283;%%
\bibitem [{\citenamefont {Hassan}\ and\ \citenamefont
  {Rosen}(2012{\natexlab{c}})}]{Hassan:2011zd}%
  \BibitemOpen
  \bibfield  {author} {\bibinfo {author} {\bibfnamefont {S.}~\bibnamefont
  {Hassan}}\ and\ \bibinfo {author} {\bibfnamefont {R.~A.}\ \bibnamefont
  {Rosen}},\ }\href {\doibase 10.1007/JHEP02(2012)126} {\bibfield  {journal}
  {\bibinfo  {journal} {JHEP}\ }\textbf {\bibinfo {volume} {1202}},\ \bibinfo
  {pages} {126} (\bibinfo {year} {2012}{\natexlab{c}})},\ \Eprint
  {http://arxiv.org/abs/1109.3515} {arXiv:1109.3515 [hep-th]} \BibitemShut
  {NoStop}%
%%CITATION = ARXIV:1109.3515;%%
\bibitem [{\citenamefont {Hinterbichler}\ and\ \citenamefont
  {Rosen}(2012)}]{Hinterbichler:2012cn}%
  \BibitemOpen
  \bibfield  {author} {\bibinfo {author} {\bibfnamefont {K.}~\bibnamefont
  {Hinterbichler}}\ and\ \bibinfo {author} {\bibfnamefont {R.~A.}\ \bibnamefont
  {Rosen}},\ }\href {\doibase 10.1007/JHEP07(2012)047} {\bibfield  {journal}
  {\bibinfo  {journal} {JHEP}\ }\textbf {\bibinfo {volume} {07}},\ \bibinfo
  {pages} {047} (\bibinfo {year} {2012})},\ \Eprint
  {http://arxiv.org/abs/1203.5783} {arXiv:1203.5783 [hep-th]} \BibitemShut
  {NoStop}%
%%CITATION = ARXIV:1203.5783;%%
\bibitem [{\citenamefont {Hinterbichler}(2012)}]{Hinterbichler:2011tt}%
  \BibitemOpen
  \bibfield  {author} {\bibinfo {author} {\bibfnamefont {K.}~\bibnamefont
  {Hinterbichler}},\ }\href {\doibase 10.1103/RevModPhys.84.671} {\bibfield
  {journal} {\bibinfo  {journal} {Rev. Mod. Phys.}\ }\textbf {\bibinfo {volume}
  {84}},\ \bibinfo {pages} {671} (\bibinfo {year} {2012})},\ \Eprint
  {http://arxiv.org/abs/1105.3735} {arXiv:1105.3735 [hep-th]} \BibitemShut
  {NoStop}%
%%CITATION = ARXIV:1105.3735;%%
\bibitem [{\citenamefont {de~Rham}(2014)}]{deRham:2014zqa}%
  \BibitemOpen
  \bibfield  {author} {\bibinfo {author} {\bibfnamefont {C.}~\bibnamefont
  {de~Rham}},\ }\href {\doibase 10.12942/lrr-2014-7} {\bibfield  {journal}
  {\bibinfo  {journal} {Living Rev. Rel.}\ }\textbf {\bibinfo {volume} {17}},\
  \bibinfo {pages} {7} (\bibinfo {year} {2014})},\ \Eprint
  {http://arxiv.org/abs/1401.4173} {arXiv:1401.4173 [hep-th]} \BibitemShut
  {NoStop}%
%%CITATION = ARXIV:1401.4173;%%
\bibitem [{\citenamefont {Solomon}(2015)}]{Solomon:2015hja}%
  \BibitemOpen
  \bibfield  {author} {\bibinfo {author} {\bibfnamefont {A.~R.}\ \bibnamefont
  {Solomon}},\ }\emph {\bibinfo {title} {{Cosmology Beyond Einstein}}},\ \href
  {\doibase 10.1007/978-3-319-46621-7} {Ph.D. thesis},\ \bibinfo  {school}
  {Cambridge U.}, \bibinfo {address} {Cham} (\bibinfo {year} {2015}),\ \Eprint
  {http://arxiv.org/abs/1508.06859} {arXiv:1508.06859 [gr-qc]} \BibitemShut
  {NoStop}%
%%CITATION = ARXIV:1508.06859;%%
\bibitem [{\citenamefont {Schmidt-May}\ and\ \citenamefont {von
  Strauss}(2016)}]{Schmidt-May:2015vnx}%
  \BibitemOpen
  \bibfield  {author} {\bibinfo {author} {\bibfnamefont {A.}~\bibnamefont
  {Schmidt-May}}\ and\ \bibinfo {author} {\bibfnamefont {M.}~\bibnamefont {von
  Strauss}},\ }\href {\doibase 10.1088/1751-8113/49/18/183001} {\bibfield
  {journal} {\bibinfo  {journal} {J. Phys.}\ }\textbf {\bibinfo {volume}
  {A49}},\ \bibinfo {pages} {183001} (\bibinfo {year} {2016})},\ \Eprint
  {http://arxiv.org/abs/1512.00021} {arXiv:1512.00021 [hep-th]} \BibitemShut
  {NoStop}%
%%CITATION = ARXIV:1512.00021;%%
\bibitem [{\citenamefont {Lim}\ \emph {et~al.}(2010)\citenamefont {Lim},
  \citenamefont {Sawicki},\ and\ \citenamefont {Vikman}}]{Lim:2010yk}%
  \BibitemOpen
  \bibfield  {author} {\bibinfo {author} {\bibfnamefont {E.~A.}\ \bibnamefont
  {Lim}}, \bibinfo {author} {\bibfnamefont {I.}~\bibnamefont {Sawicki}}, \ and\
  \bibinfo {author} {\bibfnamefont {A.}~\bibnamefont {Vikman}},\ }\href
  {\doibase 10.1088/1475-7516/2010/05/012} {\bibfield  {journal} {\bibinfo
  {journal} {JCAP}\ }\textbf {\bibinfo {volume} {1005}},\ \bibinfo {pages}
  {012} (\bibinfo {year} {2010})},\ \Eprint {http://arxiv.org/abs/1003.5751}
  {arXiv:1003.5751 [astro-ph.CO]} \BibitemShut {NoStop}%
%%CITATION = ARXIV:1003.5751;%%
\bibitem [{\citenamefont {Mirzagholi}\ and\ \citenamefont
  {Vikman}(2015)}]{Mirzagholi:2014ifa}%
  \BibitemOpen
  \bibfield  {author} {\bibinfo {author} {\bibfnamefont {L.}~\bibnamefont
  {Mirzagholi}}\ and\ \bibinfo {author} {\bibfnamefont {A.}~\bibnamefont
  {Vikman}},\ }\href {\doibase 10.1088/1475-7516/2015/06/028} {\bibfield
  {journal} {\bibinfo  {journal} {JCAP}\ }\textbf {\bibinfo {volume} {1506}},\
  \bibinfo {pages} {028} (\bibinfo {year} {2015})},\ \Eprint
  {http://arxiv.org/abs/1412.7136} {arXiv:1412.7136 [gr-qc]} \BibitemShut
  {NoStop}%
%%CITATION = ARXIV:1412.7136;%%
\bibitem [{\citenamefont {van Dam}\ and\ \citenamefont
  {Veltman}(1970)}]{vanDam:1970vg}%
  \BibitemOpen
  \bibfield  {author} {\bibinfo {author} {\bibfnamefont {H.}~\bibnamefont {van
  Dam}}\ and\ \bibinfo {author} {\bibfnamefont {M.~J.~G.}\ \bibnamefont
  {Veltman}},\ }\href {\doibase 10.1016/0550-3213(70)90416-5} {\bibfield
  {journal} {\bibinfo  {journal} {Nucl. Phys.}\ }\textbf {\bibinfo {volume}
  {B22}},\ \bibinfo {pages} {397} (\bibinfo {year} {1970})}\BibitemShut
  {NoStop}%
%%CITATION = NUPHA,B22,397;%%
\bibitem [{\citenamefont {Zakharov}(1970)}]{Zakharov:1970cc}%
  \BibitemOpen
  \bibfield  {author} {\bibinfo {author} {\bibfnamefont {V.~I.}\ \bibnamefont
  {Zakharov}},\ }\href@noop {} {\bibfield  {journal} {\bibinfo  {journal} {JETP
  Lett.}\ }\textbf {\bibinfo {volume} {12}},\ \bibinfo {pages} {312} (\bibinfo
  {year} {1970})},\ \bibinfo {note} {[Pisma Zh. Eksp. Teor.
  Fiz.12,447(1970)]}\BibitemShut {NoStop}%
%%CITATION = JTPLA,12,312;%%
\bibitem [{\citenamefont {Higuchi}(1987)}]{Higuchi:1986py}%
  \BibitemOpen
  \bibfield  {author} {\bibinfo {author} {\bibfnamefont {A.}~\bibnamefont
  {Higuchi}},\ }\href {\doibase 10.1016/0550-3213(87)90691-2} {\bibfield
  {journal} {\bibinfo  {journal} {Nucl.Phys.}\ }\textbf {\bibinfo {volume}
  {B282}},\ \bibinfo {pages} {397} (\bibinfo {year} {1987})}\BibitemShut
  {NoStop}%
%%CITATION = NUPHA,B282,397;%%
\bibitem [{\citenamefont {Vainshtein}(1972)}]{Vainshtein:1972sx}%
  \BibitemOpen
  \bibfield  {author} {\bibinfo {author} {\bibfnamefont {A.}~\bibnamefont
  {Vainshtein}},\ }\href {\doibase 10.1016/0370-2693(72)90147-5} {\bibfield
  {journal} {\bibinfo  {journal} {Phys.Lett.}\ }\textbf {\bibinfo {volume}
  {B39}},\ \bibinfo {pages} {393} (\bibinfo {year} {1972})}\BibitemShut
  {NoStop}%
%%CITATION = PHLTA,B39,393;%%
\bibitem [{\citenamefont {Babichev}\ and\ \citenamefont
  {Deffayet}(2013)}]{Babichev:2013usa}%
  \BibitemOpen
  \bibfield  {author} {\bibinfo {author} {\bibfnamefont {E.}~\bibnamefont
  {Babichev}}\ and\ \bibinfo {author} {\bibfnamefont {C.}~\bibnamefont
  {Deffayet}},\ }\href {\doibase 10.1088/0264-9381/30/18/184001} {\bibfield
  {journal} {\bibinfo  {journal} {Class.Quant.Grav.}\ }\textbf {\bibinfo
  {volume} {30}},\ \bibinfo {pages} {184001} (\bibinfo {year} {2013})},\
  \Eprint {http://arxiv.org/abs/1304.7240} {arXiv:1304.7240 [gr-qc]}
  \BibitemShut {NoStop}%
%%CITATION = ARXIV:1304.7240;%%
\bibitem [{\citenamefont {Dubovsky}(2004)}]{Dubovsky:2004sg}%
  \BibitemOpen
  \bibfield  {author} {\bibinfo {author} {\bibfnamefont {S.~L.}\ \bibnamefont
  {Dubovsky}},\ }\href {\doibase 10.1088/1126-6708/2004/10/076} {\bibfield
  {journal} {\bibinfo  {journal} {JHEP}\ }\textbf {\bibinfo {volume} {10}},\
  \bibinfo {pages} {076} (\bibinfo {year} {2004})},\ \Eprint
  {http://arxiv.org/abs/hep-th/0409124} {arXiv:hep-th/0409124 [hep-th]}
  \BibitemShut {NoStop}%
%%CITATION = HEP-TH/0409124;%%
\bibitem [{\citenamefont {Blas}\ \emph {et~al.}(2009)\citenamefont {Blas},
  \citenamefont {Comelli}, \citenamefont {Nesti},\ and\ \citenamefont
  {Pilo}}]{Blas:2009my}%
  \BibitemOpen
  \bibfield  {author} {\bibinfo {author} {\bibfnamefont {D.}~\bibnamefont
  {Blas}}, \bibinfo {author} {\bibfnamefont {D.}~\bibnamefont {Comelli}},
  \bibinfo {author} {\bibfnamefont {F.}~\bibnamefont {Nesti}}, \ and\ \bibinfo
  {author} {\bibfnamefont {L.}~\bibnamefont {Pilo}},\ }\href {\doibase
  10.1103/PhysRevD.80.044025} {\bibfield  {journal} {\bibinfo  {journal} {Phys.
  Rev.}\ }\textbf {\bibinfo {volume} {D80}},\ \bibinfo {pages} {044025}
  (\bibinfo {year} {2009})},\ \Eprint {http://arxiv.org/abs/0905.1699}
  {arXiv:0905.1699 [hep-th]} \BibitemShut {NoStop}%
%%CITATION = ARXIV:0905.1699;%%
\bibitem [{\citenamefont {Lagos}\ \emph {et~al.}(2014)\citenamefont {Lagos},
  \citenamefont {Ba{\~{n}}ados}, \citenamefont {Ferreira},\ and\ \citenamefont
  {Garc{\'i}a-S{\'a}enz}}]{Lagos:2013aua}%
  \BibitemOpen
  \bibfield  {author} {\bibinfo {author} {\bibfnamefont {M.}~\bibnamefont
  {Lagos}}, \bibinfo {author} {\bibfnamefont {M.}~\bibnamefont
  {Ba{\~{n}}ados}}, \bibinfo {author} {\bibfnamefont {P.~G.}\ \bibnamefont
  {Ferreira}}, \ and\ \bibinfo {author} {\bibfnamefont {S.}~\bibnamefont
  {Garc{\'i}a-S{\'a}enz}},\ }\href {\doibase 10.1103/PhysRevD.89.024034}
  {\bibfield  {journal} {\bibinfo  {journal} {Phys. Rev.}\ }\textbf {\bibinfo
  {volume} {D89}},\ \bibinfo {pages} {024034} (\bibinfo {year} {2014})},\
  \Eprint {http://arxiv.org/abs/1311.3828} {arXiv:1311.3828 [gr-qc]}
  \BibitemShut {NoStop}%
%%CITATION = ARXIV:1311.3828;%%
\bibitem [{\citenamefont {Motohashi}\ \emph {et~al.}(2016)\citenamefont
  {Motohashi}, \citenamefont {Suyama},\ and\ \citenamefont
  {Takahashi}}]{Motohashi:2016prk}%
  \BibitemOpen
  \bibfield  {author} {\bibinfo {author} {\bibfnamefont {H.}~\bibnamefont
  {Motohashi}}, \bibinfo {author} {\bibfnamefont {T.}~\bibnamefont {Suyama}}, \
  and\ \bibinfo {author} {\bibfnamefont {K.}~\bibnamefont {Takahashi}},\ }\href
  {\doibase 10.1103/PhysRevD.94.124021} {\bibfield  {journal} {\bibinfo
  {journal} {Phys. Rev.}\ }\textbf {\bibinfo {volume} {D94}},\ \bibinfo {pages}
  {124021} (\bibinfo {year} {2016})},\ \Eprint
  {http://arxiv.org/abs/1608.00071} {arXiv:1608.00071 [gr-qc]} \BibitemShut
  {NoStop}%
%%CITATION = ARXIV:1608.00071;%%
\bibitem [{\citenamefont {Dubovsky}\ \emph {et~al.}(2006)\citenamefont
  {Dubovsky}, \citenamefont {Gregoire}, \citenamefont {Nicolis},\ and\
  \citenamefont {Rattazzi}}]{Dubovsky:2005xd}%
  \BibitemOpen
  \bibfield  {author} {\bibinfo {author} {\bibfnamefont {S.}~\bibnamefont
  {Dubovsky}}, \bibinfo {author} {\bibfnamefont {T.}~\bibnamefont {Gregoire}},
  \bibinfo {author} {\bibfnamefont {A.}~\bibnamefont {Nicolis}}, \ and\
  \bibinfo {author} {\bibfnamefont {R.}~\bibnamefont {Rattazzi}},\ }\href
  {\doibase 10.1088/1126-6708/2006/03/025} {\bibfield  {journal} {\bibinfo
  {journal} {JHEP}\ }\textbf {\bibinfo {volume} {03}},\ \bibinfo {pages} {025}
  (\bibinfo {year} {2006})},\ \Eprint {http://arxiv.org/abs/hep-th/0512260}
  {arXiv:hep-th/0512260 [hep-th]} \BibitemShut {NoStop}%
%%CITATION = HEP-TH/0512260;%%
\bibitem [{\citenamefont {Endlich}\ \emph {et~al.}(2011)\citenamefont
  {Endlich}, \citenamefont {Nicolis}, \citenamefont {Rattazzi},\ and\
  \citenamefont {Wang}}]{Endlich:2010hf}%
  \BibitemOpen
  \bibfield  {author} {\bibinfo {author} {\bibfnamefont {S.}~\bibnamefont
  {Endlich}}, \bibinfo {author} {\bibfnamefont {A.}~\bibnamefont {Nicolis}},
  \bibinfo {author} {\bibfnamefont {R.}~\bibnamefont {Rattazzi}}, \ and\
  \bibinfo {author} {\bibfnamefont {J.}~\bibnamefont {Wang}},\ }\href {\doibase
  10.1007/JHEP04(2011)102} {\bibfield  {journal} {\bibinfo  {journal} {JHEP}\
  }\textbf {\bibinfo {volume} {04}},\ \bibinfo {pages} {102} (\bibinfo {year}
  {2011})},\ \Eprint {http://arxiv.org/abs/1011.6396} {arXiv:1011.6396
  [hep-th]} \BibitemShut {NoStop}%
%%CITATION = ARXIV:1011.6396;%%
\bibitem [{\citenamefont {Dvali}\ \emph {et~al.}(2000)\citenamefont {Dvali},
  \citenamefont {Gabadadze},\ and\ \citenamefont {Porrati}}]{Dvali:2000hr}%
  \BibitemOpen
  \bibfield  {author} {\bibinfo {author} {\bibfnamefont {G.~R.}\ \bibnamefont
  {Dvali}}, \bibinfo {author} {\bibfnamefont {G.}~\bibnamefont {Gabadadze}}, \
  and\ \bibinfo {author} {\bibfnamefont {M.}~\bibnamefont {Porrati}},\ }\href
  {\doibase 10.1016/S0370-2693(00)00669-9} {\bibfield  {journal} {\bibinfo
  {journal} {Phys. Lett.}\ }\textbf {\bibinfo {volume} {B485}},\ \bibinfo
  {pages} {208} (\bibinfo {year} {2000})},\ \Eprint
  {http://arxiv.org/abs/hep-th/0005016} {arXiv:hep-th/0005016 [hep-th]}
  \BibitemShut {NoStop}%
%%CITATION = HEP-TH/0005016;%%
\bibitem [{\citenamefont {Deffayet}(2001)}]{Deffayet:2000uy}%
  \BibitemOpen
  \bibfield  {author} {\bibinfo {author} {\bibfnamefont {C.}~\bibnamefont
  {Deffayet}},\ }\href {\doibase 10.1016/S0370-2693(01)00160-5} {\bibfield
  {journal} {\bibinfo  {journal} {Phys. Lett.}\ }\textbf {\bibinfo {volume}
  {B502}},\ \bibinfo {pages} {199} (\bibinfo {year} {2001})},\ \Eprint
  {http://arxiv.org/abs/hep-th/0010186} {arXiv:hep-th/0010186 [hep-th]}
  \BibitemShut {NoStop}%
%%CITATION = HEP-TH/0010186;%%
\bibitem [{\citenamefont {Deffayet}\ \emph {et~al.}(2002)\citenamefont
  {Deffayet}, \citenamefont {Dvali},\ and\ \citenamefont
  {Gabadadze}}]{Deffayet:2001pu}%
  \BibitemOpen
  \bibfield  {author} {\bibinfo {author} {\bibfnamefont {C.}~\bibnamefont
  {Deffayet}}, \bibinfo {author} {\bibfnamefont {G.~R.}\ \bibnamefont {Dvali}},
  \ and\ \bibinfo {author} {\bibfnamefont {G.}~\bibnamefont {Gabadadze}},\
  }\href {\doibase 10.1103/PhysRevD.65.044023} {\bibfield  {journal} {\bibinfo
  {journal} {Phys. Rev.}\ }\textbf {\bibinfo {volume} {D65}},\ \bibinfo {pages}
  {044023} (\bibinfo {year} {2002})},\ \Eprint
  {http://arxiv.org/abs/astro-ph/0105068} {arXiv:astro-ph/0105068 [astro-ph]}
  \BibitemShut {NoStop}%
%%CITATION = ASTRO-PH/0105068;%%
\bibitem [{\citenamefont {Gorbunov}\ \emph {et~al.}(2006)\citenamefont
  {Gorbunov}, \citenamefont {Koyama},\ and\ \citenamefont
  {Sibiryakov}}]{Gorbunov:2005zk}%
  \BibitemOpen
  \bibfield  {author} {\bibinfo {author} {\bibfnamefont {D.}~\bibnamefont
  {Gorbunov}}, \bibinfo {author} {\bibfnamefont {K.}~\bibnamefont {Koyama}}, \
  and\ \bibinfo {author} {\bibfnamefont {S.}~\bibnamefont {Sibiryakov}},\
  }\href {\doibase 10.1103/PhysRevD.73.044016} {\bibfield  {journal} {\bibinfo
  {journal} {Phys. Rev.}\ }\textbf {\bibinfo {volume} {D73}},\ \bibinfo {pages}
  {044016} (\bibinfo {year} {2006})},\ \Eprint
  {http://arxiv.org/abs/hep-th/0512097} {arXiv:hep-th/0512097 [hep-th]}
  \BibitemShut {NoStop}%
%%CITATION = HEP-TH/0512097;%%
\bibitem [{\citenamefont {Charmousis}\ \emph {et~al.}(2006)\citenamefont
  {Charmousis}, \citenamefont {Gregory}, \citenamefont {Kaloper},\ and\
  \citenamefont {Padilla}}]{Charmousis:2006pn}%
  \BibitemOpen
  \bibfield  {author} {\bibinfo {author} {\bibfnamefont {C.}~\bibnamefont
  {Charmousis}}, \bibinfo {author} {\bibfnamefont {R.}~\bibnamefont {Gregory}},
  \bibinfo {author} {\bibfnamefont {N.}~\bibnamefont {Kaloper}}, \ and\
  \bibinfo {author} {\bibfnamefont {A.}~\bibnamefont {Padilla}},\ }\href
  {\doibase 10.1088/1126-6708/2006/10/066} {\bibfield  {journal} {\bibinfo
  {journal} {JHEP}\ }\textbf {\bibinfo {volume} {10}},\ \bibinfo {pages} {066}
  (\bibinfo {year} {2006})},\ \Eprint {http://arxiv.org/abs/hep-th/0604086}
  {arXiv:hep-th/0604086 [hep-th]} \BibitemShut {NoStop}%
%%CITATION = HEP-TH/0604086;%%
\bibitem [{\citenamefont {Khoury}\ \emph {et~al.}(2018)\citenamefont {Khoury},
  \citenamefont {Sakstein},\ and\ \citenamefont {Solomon}}]{Khoury:2018vdv}%
  \BibitemOpen
  \bibfield  {author} {\bibinfo {author} {\bibfnamefont {J.}~\bibnamefont
  {Khoury}}, \bibinfo {author} {\bibfnamefont {J.}~\bibnamefont {Sakstein}}, \
  and\ \bibinfo {author} {\bibfnamefont {A.~R.}\ \bibnamefont {Solomon}},\
  }\href {\doibase 10.1088/1475-7516/2018/08/024} {\bibfield  {journal}
  {\bibinfo  {journal} {JCAP}\ }\textbf {\bibinfo {volume} {1808}},\ \bibinfo
  {pages} {024} (\bibinfo {year} {2018})},\ \Eprint
  {http://arxiv.org/abs/1805.05937} {arXiv:1805.05937 [hep-th]} \BibitemShut
  {NoStop}%
%%CITATION = ARXIV:1805.05937;%%
\bibitem [{\citenamefont {Aghanim}\ \emph {et~al.}(2018)\citenamefont {Aghanim}
  \emph {et~al.}}]{Aghanim:2018eyx}%
  \BibitemOpen
  \bibfield  {author} {\bibinfo {author} {\bibfnamefont {N.}~\bibnamefont
  {Aghanim}} \emph {et~al.} (\bibinfo {collaboration} {Planck}),\ }\href@noop
  {} {\  (\bibinfo {year} {2018})},\ \Eprint {http://arxiv.org/abs/1807.06209}
  {arXiv:1807.06209 [astro-ph.CO]} \BibitemShut {NoStop}%
%%CITATION = ARXIV:1807.06209;%%
\bibitem [{\citenamefont {Chen}\ \emph {et~al.}(2001)\citenamefont {Chen},
  \citenamefont {Scherrer},\ and\ \citenamefont {Steigman}}]{Chen:2000xxa}%
  \BibitemOpen
  \bibfield  {author} {\bibinfo {author} {\bibfnamefont {X.-l.}\ \bibnamefont
  {Chen}}, \bibinfo {author} {\bibfnamefont {R.~J.}\ \bibnamefont {Scherrer}},
  \ and\ \bibinfo {author} {\bibfnamefont {G.}~\bibnamefont {Steigman}},\
  }\href {\doibase 10.1103/PhysRevD.63.123504} {\bibfield  {journal} {\bibinfo
  {journal} {Phys. Rev.}\ }\textbf {\bibinfo {volume} {D63}},\ \bibinfo {pages}
  {123504} (\bibinfo {year} {2001})},\ \Eprint
  {http://arxiv.org/abs/astro-ph/0011531} {arXiv:astro-ph/0011531 [astro-ph]}
  \BibitemShut {NoStop}%
%%CITATION = ASTRO-PH/0011531;%%
\bibitem [{\citenamefont {Cyburt}\ \emph {et~al.}(2016)\citenamefont {Cyburt},
  \citenamefont {Fields}, \citenamefont {Olive},\ and\ \citenamefont
  {Yeh}}]{Cyburt:2015mya}%
  \BibitemOpen
  \bibfield  {author} {\bibinfo {author} {\bibfnamefont {R.~H.}\ \bibnamefont
  {Cyburt}}, \bibinfo {author} {\bibfnamefont {B.~D.}\ \bibnamefont {Fields}},
  \bibinfo {author} {\bibfnamefont {K.~A.}\ \bibnamefont {Olive}}, \ and\
  \bibinfo {author} {\bibfnamefont {T.-H.}\ \bibnamefont {Yeh}},\ }\href
  {\doibase 10.1103/RevModPhys.88.015004} {\bibfield  {journal} {\bibinfo
  {journal} {Rev. Mod. Phys.}\ }\textbf {\bibinfo {volume} {88}},\ \bibinfo
  {pages} {015004} (\bibinfo {year} {2016})},\ \Eprint
  {http://arxiv.org/abs/1505.01076} {arXiv:1505.01076 [astro-ph.CO]}
  \BibitemShut {NoStop}%
%%CITATION = ARXIV:1505.01076;%%
\bibitem [{\citenamefont {{Fixsen}}(2009)}]{2009ApJ...707..916F}%
  \BibitemOpen
  \bibfield  {author} {\bibinfo {author} {\bibfnamefont {D.~J.}\ \bibnamefont
  {{Fixsen}}},\ }\href {\doibase 10.1088/0004-637X/707/2/916} {\bibfield
  {journal} {\bibinfo  {journal} {\apj}\ }\textbf {\bibinfo {volume} {707}},\
  \bibinfo {pages} {916} (\bibinfo {year} {2009})},\ \Eprint
  {http://arxiv.org/abs/0911.1955} {arXiv:0911.1955 [astro-ph.CO]} \BibitemShut
  {NoStop}%
\bibitem [{\citenamefont {Akrami}\ \emph {et~al.}(2015)\citenamefont {Akrami},
  \citenamefont {Hassan}, \citenamefont {K{\"{o}}nnig}, \citenamefont
  {Schmidt-May},\ and\ \citenamefont {Solomon}}]{Akrami:2015qga}%
  \BibitemOpen
  \bibfield  {author} {\bibinfo {author} {\bibfnamefont {Y.}~\bibnamefont
  {Akrami}}, \bibinfo {author} {\bibfnamefont {S.~F.}\ \bibnamefont {Hassan}},
  \bibinfo {author} {\bibfnamefont {F.}~\bibnamefont {K{\"{o}}nnig}}, \bibinfo
  {author} {\bibfnamefont {A.}~\bibnamefont {Schmidt-May}}, \ and\ \bibinfo
  {author} {\bibfnamefont {A.~R.}\ \bibnamefont {Solomon}},\ }\href {\doibase
  10.1016/j.physletb.2015.06.062} {\bibfield  {journal} {\bibinfo  {journal}
  {Phys. Lett.}\ }\textbf {\bibinfo {volume} {B748}},\ \bibinfo {pages} {37}
  (\bibinfo {year} {2015})},\ \Eprint {http://arxiv.org/abs/1503.07521}
  {arXiv:1503.07521 [gr-qc]} \BibitemShut {NoStop}%
%%CITATION = ARXIV:1503.07521;%%
\bibitem [{\citenamefont {Abbott}\ \emph {et~al.}(2017)\citenamefont {Abbott}
  \emph {et~al.}}]{Abbott:2017vtc}%
  \BibitemOpen
  \bibfield  {author} {\bibinfo {author} {\bibfnamefont {B.~P.}\ \bibnamefont
  {Abbott}} \emph {et~al.} (\bibinfo {collaboration} {LIGO Scientific,
  VIRGO}),\ }\href {\doibase 10.1103/PhysRevLett.118.221101,
  10.1103/PhysRevLett.121.129901} {\bibfield  {journal} {\bibinfo  {journal}
  {Phys. Rev. Lett.}\ }\textbf {\bibinfo {volume} {118}},\ \bibinfo {pages}
  {221101} (\bibinfo {year} {2017})},\ \bibinfo {note} {[Erratum: Phys. Rev.
  Lett.121,no.12,129901(2018)]},\ \Eprint {http://arxiv.org/abs/1706.01812}
  {arXiv:1706.01812 [gr-qc]} \BibitemShut {NoStop}%
%%CITATION = ARXIV:1706.01812;%%
\bibitem [{\citenamefont {de~Rham}\ \emph {et~al.}(2017)\citenamefont
  {de~Rham}, \citenamefont {Deskins}, \citenamefont {Tolley},\ and\
  \citenamefont {Zhou}}]{deRham:2016nuf}%
  \BibitemOpen
  \bibfield  {author} {\bibinfo {author} {\bibfnamefont {C.}~\bibnamefont
  {de~Rham}}, \bibinfo {author} {\bibfnamefont {J.~T.}\ \bibnamefont
  {Deskins}}, \bibinfo {author} {\bibfnamefont {A.~J.}\ \bibnamefont {Tolley}},
  \ and\ \bibinfo {author} {\bibfnamefont {S.-Y.}\ \bibnamefont {Zhou}},\
  }\href {\doibase 10.1103/RevModPhys.89.025004} {\bibfield  {journal}
  {\bibinfo  {journal} {Rev. Mod. Phys.}\ }\textbf {\bibinfo {volume} {89}},\
  \bibinfo {pages} {025004} (\bibinfo {year} {2017})},\ \Eprint
  {http://arxiv.org/abs/1606.08462} {arXiv:1606.08462 [astro-ph.CO]}
  \BibitemShut {NoStop}%
%%CITATION = ARXIV:1606.08462;%%
\bibitem [{\citenamefont {Enander}\ \emph {et~al.}(2015)\citenamefont
  {Enander}, \citenamefont {Akrami}, \citenamefont {M{\"o}rtsell},
  \citenamefont {Renneby},\ and\ \citenamefont {Solomon}}]{Enander:2015vja}%
  \BibitemOpen
  \bibfield  {author} {\bibinfo {author} {\bibfnamefont {J.}~\bibnamefont
  {Enander}}, \bibinfo {author} {\bibfnamefont {Y.}~\bibnamefont {Akrami}},
  \bibinfo {author} {\bibfnamefont {E.}~\bibnamefont {M{\"o}rtsell}}, \bibinfo
  {author} {\bibfnamefont {M.}~\bibnamefont {Renneby}}, \ and\ \bibinfo
  {author} {\bibfnamefont {A.~R.}\ \bibnamefont {Solomon}},\ }\href {\doibase
  10.1103/PhysRevD.91.084046} {\bibfield  {journal} {\bibinfo  {journal} {Phys.
  Rev.}\ }\textbf {\bibinfo {volume} {D91}},\ \bibinfo {pages} {084046}
  (\bibinfo {year} {2015})},\ \Eprint {http://arxiv.org/abs/1501.02140}
  {arXiv:1501.02140 [astro-ph.CO]} \BibitemShut {NoStop}%
%%CITATION = ARXIV:1501.02140;%%
\end{thebibliography}%

\end{document}